\documentclass[journal, draftcls, a4paper, onecolumn]{IEEEtran}

\usepackage{cite}

\usepackage{graphicx}

\usepackage[cmex10]{amsmath}
\usepackage{amssymb}
\DeclareMathOperator{\tr}{tr}
\DeclareMathOperator{\cov}{cov}

\usepackage{xcolor}

\usepackage[tight,footnotesize]{subfigure}

\allowdisplaybreaks

\begin{document}

\definecolor{red}{rgb}{1,0,0}

%
\title{The Sum Rate of Vector Gaussian Multiple Description Coding with Tree-Structured Covariance Distortion Constraints}


\author{Yinfei~Xu,~\IEEEmembership{Student Member,~IEEE,}
   ~Jun~Chen,~\IEEEmembership{Senior Member,~IEEE,}
   ~and~Qiao~Wang,~\IEEEmembership{Member,~IEEE}

\thanks{
This paper was presented in part at the 2015 IEEE International Symposium on Information Theory.

Yinfei Xu was with the School of Information Science and Engineering, Southeast University, Nanjing 210096, China. He is now with the Institute of Network Coding, The Chinese University of Hong Kong, Hong Kong (email: yinfeixu@seu.edu.cn).

Jun Chen is with the Department of Electrical and Computer Engineering, McMaster University, Hamilton, ON L8S 4K1, Canada (email: junchen@ece.mcmaster.ca).

Qiao Wang is with the School of Information Science and Engineering, Southeast University, Nanjing 210096, China (email: qiaowang@seu.edu.cn).
}

}

\date{\today}
\maketitle

\begin{abstract}
A single-letter lower bound on the sum rate of multiple description coding with tree-structured distortion constraints is established by generalizing Ozarow's celebrated converse argument through the introduction of auxiliary random variables that form a Markov tree. For the quadratic vector Gaussian case, this lower bound is shown to be achievable by an extended version of the El Gamal-Cover scheme, yielding a complete sum-rate characterization.



\end{abstract}


\newcounter{mytempeqncnt}
\newtheorem{theorem}{Theorem}
\newtheorem{lemma}{Lemma}
\newtheorem{definition}{Definition}
\newtheorem{remark}{Remark}

\section{Introduction}
In multiple description coding, a source is encoded into $M$ descriptions with rates $R_1,\cdots,R_M$, respectively, such that each non-empty subset of these descriptions can be used to produce a (possibly lossy) reconstruction of the source. The fundamental information-theoretic problem here is to characterize the rate region, i.e., the closure of the set of the admissible rate tuples $(R_1,\cdots,R_M)$, subject to the given distortion constraints on the reconstructions. Early work on this problem was mostly devoted to the two-description case. In particular, El Gamal and Cover \cite{EGC82} derived an inner bound of the two-description rate region, which was shown to be tight in the no excess sum-rate case  \cite{Al85} and in the case where one of the descriptions is required to reconstruct a deterministic function of the source \cite{FY02}; Zhang and Berger \cite{ZB87} obtained an improved inner bound of the two-description rate region by incorporating a common description layer into the El Gamal-Cover scheme. In contrast, recent years have seen extensive research on the general $M$-description case (see, e.g., \cite{VKG03,PPR04,PPR05,TC10,VAR16}). Although these investigations have revealed many interesting results and considerably deepened our understanding of the subject, a complete characterization of the rate region, even for the two-description case, is still widely considered to be out of reach.

One way to simplify the problem is to consider a special class of sources and distortion measures. In this respect, special attentions have been paid to the Gaussian source and the mean squared error distortion measure (known as the quadratic Gaussian case). In his lauded paper \cite{O80}, Ozarow proved the tightness of the El Gamal-Cover inner bound for the quadratic Gaussian two-description problem via an ingenious converse argument, in which an auxiliary random variable is introduced to exploit an implicit conditional independence structure in the El Gamal-Cover scheme. Ozarow's work ignited the hope of solving the general quadratic Gaussian $M$-description problem. Unfortunately, this task turns out to be rather formidable (if not impossible). Nevertheless, the bounding technique in \cite{O80}  has been extended to obtain conclusive results for some special cases where the reconstruction distortion constraints are only imposed on certain subsets of descriptions, including the quadratic Gaussian $M$-description problem with individual and central distortion constraints \cite{WV07} as well as the quadratic Gaussian $M$-description problem with individual and hierarchical distortion constraints \cite{C09}. We go one step further in this work by considering the more general tree-structured distortion constraints. It will been seen that Ozarow's converse argument admits a natural generalization in this context, leading to a complete sum-rate characterization.

The importance of conditional independence structures in the converse arguments is widely recognized, especially for the distributed source coding problems (see, e.g., \cite{Oohama98,Oohama05,WA08,WTV08,TVW10,WCW10,Wang13,Wang14,Oohama14}). It should be noted that, in distributed source coding, the conditional independence structures,
which are either directly present in the source models or created through the introduction of auxiliary random variables, do not depend on the adopted schemes. In contrast, due to its centralized encoding nature, no such non-trivial scheme-independent conditional independence structures exist in multiple description coding. Indeed, Ozarow's converse argument is tailored to an implicit conditional independence structure which is specific to the El Gamal-Cover scheme. In this work we extend Ozarow's method to cope with more sophisticated conditional independence structures in a generalized El Gamal-Cover scheme optimized for tree-structured distortion constraints.

The remainder of this paper is organized as follows. The problem formulation and the main results are stated in Section \ref{sec:formulation}. Section \ref{sec:proof} is devoted to characterizing the minimum sum rate of vector Gaussian multiple description coding with tree-structured covariance distortion constraints. We conclude the paper in Section \ref{sec:conclusion}.

\section{Problem Formulation and Main Results}\label{sec:formulation}

\begin{figure}
\centering
\includegraphics[width=0.7\textwidth]{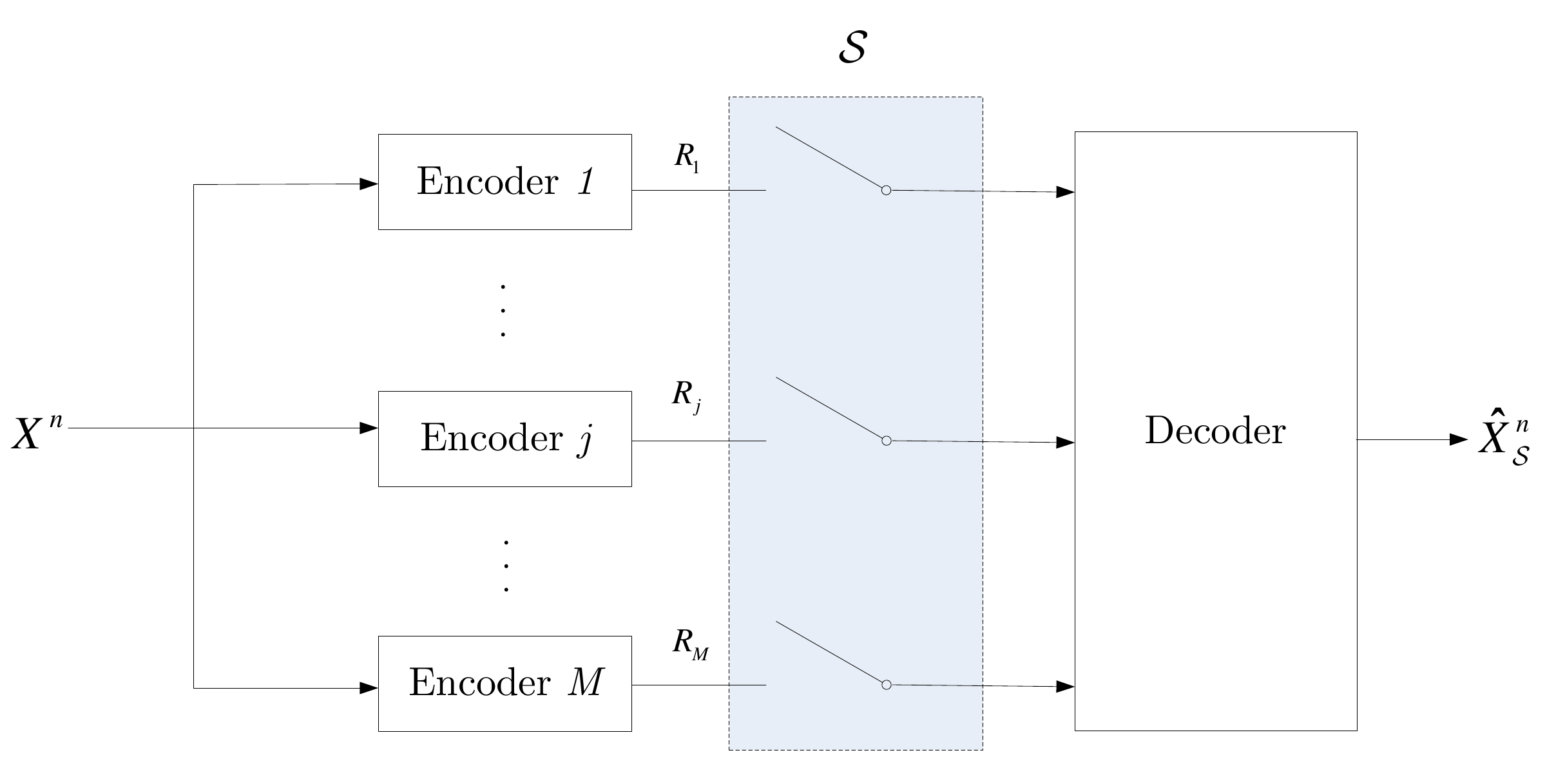}
\caption{The multiple description coding system.}
\label{fig1}
\centering
\end{figure}

Consider a multiple description coding system (see Fig. \ref{fig1}) with $M$ ($M\geq 2$) encoders, each generating a description of the source, and a decoder which produces a reconstruction of the source based on the received descriptions. We assume that the source is an i.i.d. process $\{ X(t)\}_{t=1}^{\infty}$ with marginal distribution $p(x)$ over alphabet $\mathcal{X}$,
and let
\begin{align*}
d: \mathcal{X} \times \mathcal{\hat{X}} \rightarrow [0, \infty]
\end{align*}
be a distortion measure, where $\hat{\mathcal{X}}$ is the reconstruction alphabet.

\begin{definition}
Let $\mathcal{T}$ be a collection of nonempty subsets of $\mathcal{I}_M\triangleq\{1,\cdots,M\}$. We say that a rate tuple $(R_1,\cdots,R_M)$ is achievable subject to distortion constraints $d_{\mathcal{S}}$, $\mathcal{S}\in\mathcal{T}$, if there exist encoding functions
\begin{align*}
\varphi_{j}^{(n)}: \mathcal{X}^{n} \rightarrow \mathcal{C}_{j}^{(n)}, \quad j=1,\cdots,M,
\end{align*}
and decoding functions
\begin{align*}
\psi_{\mathcal{S}}^{(n)}: \prod_{j\in\mathcal{S}}\mathcal{C}_{j}^{(n)} \rightarrow \hat{\mathcal{X}}^{n},\quad\mathcal{S}\in\mathcal{T},
\end{align*}
such that
\begin{align*}
& \frac{1}{n} \log  | \mathcal{C}_{j}^{(n)}| \leq R_{j}, \quad j =1,\cdots,M,  \\
& \frac{1}{n} \sum_{t=1}^{n} \mathbb{E} [ d(X(t), \hat{X}_{\mathcal{S}}(t)) ] \leq d_{\mathcal{S}}, \quad\mathcal{S}\in\mathcal{T},
\end{align*}
where $\hat{X}_{\mathcal{S}}^{n} =  \psi_{\mathcal{S}}^{(n)} (  (\varphi_{j}^{(n)}(X^{n}))_{j \in \mathcal{S}} )$. The rate region $\mathcal{R}((d_{\mathcal{S}})_{\mathcal{S}\in\mathcal{T}})$ is the closure of the set of all such achievable rate tuples, and the minimum sum rate $R((d_{\mathcal{S}})_{\mathcal{S}\in\mathcal{T}})$ is defined as
\begin{align*}
R((d_{\mathcal{S}})_{\mathcal{S}\in\mathcal{T}})=\min\limits_{(R_1,\cdots,R_M)\in\mathcal{R}((d_{\mathcal{S}})_{\mathcal{S}\in\mathcal{T}})}\sum\limits_{j=1}^MR_j.
\end{align*}
\end{definition}


Choosing $\mathcal{T}=2^{\mathcal{I}_M}_+$ gives the generic form of $M$-description coding, where $2^{\mathcal{I}_M}_+$ is the collection of all nonempty subsets of $\mathcal{I}_M$. It is clear that every other choice of $\mathcal{T}$ corresponds to a degenerate case of the generic form with $d_{\mathcal{S}}=\infty$ for $\mathcal{S}\in2^{\mathcal{I}_M}_+\backslash\mathcal{T}$. In this work we consider the case where $\mathcal{T}$ has a tree structure. Specifically, we say that $\mathcal{T}$ has a tree structure if, for any $\mathcal{S}_1,\mathcal{S}_2\in\mathcal{T}$, one of the following statements is true:
\begin{enumerate}
\item $\mathcal{S}_1\subseteq\mathcal{S}_2$,

\item $\mathcal{S}_2\subseteq\mathcal{S}_1$,

\item $\mathcal{S}_1\cap\mathcal{S}_2=\emptyset$.
\end{enumerate}
The following settings considered in the literature correspond to two special tree structures:
\begin{itemize}
\item individual and central distortion constraints \cite{WV07}, i.e., $\mathcal{T}=\{\{1\},\cdots,\{M\},\{1,\cdots,M\}\}$,

\item individual and hierarchical distortion constraints \cite{C09}, i.e., $\mathcal{T}=\{\{1\},\cdots,\{M\},\{1,2\},\cdots,\{1,\cdots,M\}\}$.
\end{itemize}
Note that $2^{\mathcal{I}_M}_+$ has a tree structure if and only if $M=2$ (assuming $M\geq 2$); in this sense, the two-description problem is inherently simpler than the general $M$-description problem.

\begin{figure}
\centering
\includegraphics[width=0.5\textwidth]{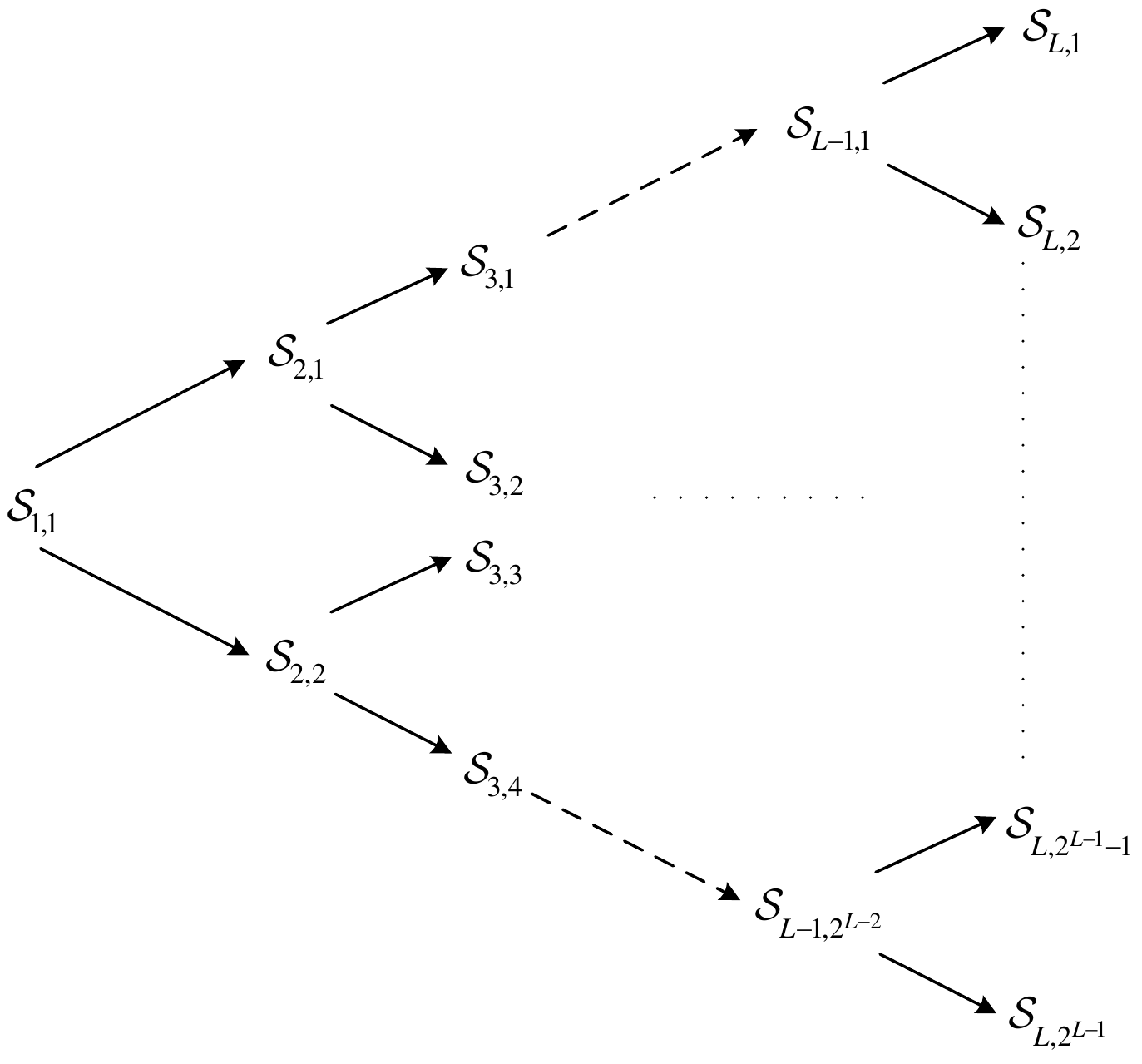}
\caption{A perfect binary tree.}
\label{fig2}
\centering
\end{figure}

For notational simplicity, henceforth we shall assume  that $M=2^{L-1}$ for some $L\geq 2$ and that $\mathcal{T}$ has a perfect binary-tree structure (see Fig. \ref{fig2}), i.e.,
\begin{align*}
\mathcal{T}=\{\mathcal{S}_{k,i}:k=1,\cdots,L; i=1,\cdots,2^{k-1}\},
\end{align*}
where
\begin{align*}
\mathcal{S}_{k,i}=\left\{ j\in\mathbb{N}:     \frac{2^{L}(i-1)}{2^k} < j \leq  \frac{2^{L}i}{2^{k}}    \right\}.
\end{align*}
It is clear that this assumption incurs no loss of generality since
every tree can be converted to a perfect binary tree via inserting dummy nodes (which corresponds to inserting dummy descriptions and imposing redundant distortion constraints in the multiple description coding system) and relabelling. 




We shall establish a single-letter lower bound on the sum rate of multiple description coding for the case where $\mathcal{T}$ has a perfect binary-tree structure.. The proof is based on the idea of augmenting the probability space with auxiliary random variables that form a Markov tree. This idea was originated in Ozarow's seminal work \cite{O80}, in which a single auxiliary random variable is introduced. Later \cite{C09,TMD09,SSC14} considered a more general construction with multiple auxiliary random variables forming a Markov chain (see also \cite{SCWL13} for a related construction). Our Markov-tree construction can be viewed as a further generalization along this line of development.


Let $\mathcal{P}$ denote the set of conditional distributions $p(\underline{z}|x)$ satisfying the binary Markov tree condition
\begin{align}
p(\underline{z}|x)= p(z_{1,1}|x) \prod_{k=1}^{L-2}  \prod_{i=1}^{2^{k-1}} p(z_{k+1, 2i-1}, z_{k+1, 2i} | z_{k,i}),\label{eq:Markovtree}
\end{align}
where $\underline{z}=(z_{k,i})_{k=1,\cdots, L-1; i=1,\cdots,2^{k-1}}$.
Moreover, let $\mathcal{P}(\underline{d})$ denote the set of conditional distributions $p(\underline{\hat{x}}|x)$ such that the induced $p(x)p(\underline{\hat{x}}|x)$ satisfies
\begin{align*}
\mathbb{E} [ d(X, \hat{X}_{\mathcal{S}_{k,i}}) ] \leq d_{\mathcal{S}_{k,i}},\quad k=1,\cdots, L; i=1,\cdots,2^{k-1},
\end{align*}
where $\underline{d}=(d_{\mathcal{S}_{k,i}})_{k=1,\cdots, L; i=1,\cdots,2^{k-1}}$ and $\underline{\hat{x}}=(\hat{x}_{\mathcal{S}_{k,i}})_{k=1,\cdots, L; i=1,\cdots,2^{k-1}}$. Let
\begin{align*}
\eta(p(\hat{x}|x),p(\underline{z}|x))&=I(X; \hat{X}_{\mathcal{S}_{1,1}}|Z_{1,1})\nonumber\\
&\quad+ \sum_{k=1}^{L-2} \sum_{i=1}^{2^{k-1}} (I(Z_{k,i}; \hat{X}_{\mathcal{S}_{k+1,2i-1}} | Z_{k+1,2i-1}) + I(Z_{k,i};  \hat{X}_{\mathcal{S}_{k+1,2i}} | Z_{k+1,2i})) \nonumber \\
 &\quad+ \sum_{i=1}^{2^{L-2}} ( I (Z_{L-1,i}; \hat{X}_{\mathcal{S}_{L ,2i-1}}) + I (Z_{L-1,i}; \hat{X}_{\mathcal{S}_{L ,2i}}))
\end{align*}
with $p(\underline{z},x,\underline{\hat{x}})=p(\underline{z}|x)p(x)p(\underline{\hat{x}}|x)$, and define
\begin{align*}
r(\underline{d})= \inf\limits_{p(\hat{x}|x)\in\mathcal{P}(\underline{d})}\sup\limits_{p(\underline{z}|x)\in\mathcal{P}}\eta(p(\hat{x}|x),p(\underline{z}|x)).
\end{align*}

\begin{theorem}\label{lemma:LB}
$R(\underline{d})\geq r(\underline{d})$.
\end{theorem}
\begin{IEEEproof}
See Appendix \ref{LB}.
\end{IEEEproof}









In this work, we focus on the case where $\{ X(t)\}_{t=1}^{\infty}$ is a stationary and memoryless process with each $X(t)$ being an $ m \times 1$ Gaussian random vector of mean zero and covariance matrix $\mathbf{\Sigma}_{X}\succ \mathbf{0}$. Moreover, we adopt the covariance distortion constraints
\begin{align*}
 \frac{1}{n} \sum_{t=1}^{n} \mathbb{E} [( X(t)-\hat{X}_{\mathcal{S}_{k,i}}(t))( X(t)-\hat{X}_{\mathcal{S}_{k,i}}(t))^T] \preceq \mathbf{D}_{\mathcal{S}_{k,i}}, \quad k=1,\cdots,L; i=1,\cdots,2^{k-1},
\end{align*}
and, without loss of generality, assume  that $ \mathbf{0} \prec \mathbf{D}_{\mathcal{S}_{k,i}} \preceq \mathbf{\Sigma}_{X}$, $k=1,\cdots,L; i=1,\cdots,2^{k-1}$. The corresponding minimum sum rate will be denoted by $R_G(\underline{\mathbf{D}})$, where $\underline{\mathbf{D}}=((\mathbf{D}_{\mathcal{S}_{k,i}})_{k=1,\cdots,L; i=1,\cdots,2^{k-1}})$.

The next result provides a computable characterization of $R_G(\underline{\mathbf{D}})$. Define
\begin{align}
R^{*}_G(\underline{\mathbf{D}})&= \max\limits_{\underline{\mathbf{\Theta}}} \frac{1}{2} \log \frac{|\mathbf{\Sigma}_{X}|}{|\mathbf{D}_{\mathcal{S}_{1,1}}|}  \nonumber \\
& \qquad + \sum_{k=1}^{L-1} \sum_{i=1}^{2^{k-1}} \frac{1}{2} \log \frac{  |\mathbf{\Sigma}_{X}| |\mathbf{D}_{\mathcal{S}_{k,i}}\mathbf{\Sigma}_{X}^{-1}(\mathbf{\Sigma}_{X} - \mathbf{\Theta}_{k,i}   ) +  \mathbf{\Theta}_{k,i} |      }{ |\mathbf{D}_{\mathcal{S}_{k+1,2i-1}}\mathbf{\Sigma}_{X}^{-1}(\mathbf{\Sigma}_{X} - \mathbf{\Theta}_{k,i}  ) +  \mathbf{\Theta}_{k,i} |   |\mathbf{D}_{\mathcal{S}_{k+1,2i}}\mathbf{\Sigma}_{X}^{-1}(\mathbf{\Sigma}_{X} - \mathbf{\Theta}_{k,i} ) +  \mathbf{\Theta}_{k,i} |       }  \nonumber \\
& \text{subject to} \quad\mathbf{\Sigma}_{X} \succeq \mathbf{\Theta}_{k+1, 2i-1}, \mathbf{\Theta}_{k+1, 2i} \succeq \mathbf{\Theta}_{k,i} \succeq \mathbf{0}, \quad k=1,\cdots,L-2; i=1,\cdots,2^{k-1}, \label{eq:main}
\end{align}
where $\underline{\mathbf{\Theta}}=(\mathbf{\Theta}_{k,i})_{k=1,\cdots,L-1; i=1,\cdots,2^{k-1}}$.

\begin{theorem}\label{thm_main}
$R_G(\underline{\mathbf{D}})=R^{*}_G(\underline{\mathbf{D}})$.
\end{theorem}
\begin{IEEEproof}
See Section \ref{sec:proof}.
\end{IEEEproof}


\section{Proof of Theorem \ref{thm_main}}\label{sec:proof}

We divide the proof of Theorem \ref{thm_main} into two parts: Section \ref{sec:converse} is devoted to establishing the converse part (i.e., $R_G(\underline{\mathbf{D}})\geq R^{*}_G(\underline{\mathbf{D}})$) while Section \ref{sec:achievability} is devoted to establishing the achievability part (i.e, $R_G(\underline{\mathbf{D}})\leq R^{*}_G(\underline{\mathbf{D}})$).

\subsection{The Converse Part}\label{sec:converse}
Translating Theorem \ref{lemma:LB} to the quadratic vector Gaussian setting gives
\begin{align*}
R_G(\underline{\mathbf{D}})\geq r_G(\underline{\mathbf{D}})\triangleq\inf\limits_{p(\hat{x}|x)\in\mathcal{P}(\underline{\mathbf{D}})}\sup\limits_{p(\underline{z}|x)\in\mathcal{P}}\eta(p(\hat{x}|x),p(\underline{z}|x)),
\end{align*}
where $\mathcal{P}(\underline{\mathbf{D}})$ denotes the set of conditional distributions $p(\underline{\hat{x}}|x)$ such that the induced $p(x)p(\underline{\hat{x}}|x)$ satisfies
\begin{align*}
\mathbb{E} [( X-\hat{X}_{\mathcal{S}_{k,i}})( X-\hat{X}_{\mathcal{S}_{k,i}})^T] \preceq \mathbf{D}_{\mathcal{S}_{k,i}}, \quad k=1,\cdots,L; i=1,\cdots,2^{k-1}.
\end{align*}
Therefore, it suffices to prove $r_G(\underline{\mathbf{D}})\geq R^{*}_G(\underline{\mathbf{D}})$.

Let $W_{k,i}$ be an $m\times 1$ Gaussian random vector of mean zero and covariance matrix $\mathbf{\Sigma}_{W_{k,i}}\succ\mathbf{0}$, $k=1,\cdots, L-1; i=1,\cdots,2^{k-1}$. We assume that $X$, $W_{k,i}$,  $k=1,\cdots, L-1; i=1,\cdots,2^{k-1}$, are mutually independent. Define
\begin{align*}
&N_{1,1}=W_{1,1},\\
&N_{k+1,2i-1}=N_{k,i}+W_{k+1,2i-1},\quad k=1,\cdots,L-2; i=1,\cdots,2^{k-1},\\
&N_{k+1,2i}=N_{k,i}+W_{k+1,2i},\quad k=1,\cdots,L-2; i=1,\cdots,2^{k-1}.
\end{align*}
The covariance matrix of $N_{k,i}$ is denoted by $\mathbf{\Sigma}_{N_{k,i}}$, $k=1,\cdots,L-1; i=1,\cdots,2^{k-1}$. Let
\begin{align*}
Z_{k,i}=X+N_{k,i},\quad k=1,\cdots,L-1; i=1,\cdots,2^{k-1}.
\end{align*}
It is clear that $\underline{Z}\triangleq(Z_{k,i})_{k=1,\cdots,L-1; i=1,\cdots,2^{k-1}}$ and $X$ form a binary Gauss-Markov tree; as a consequence, we have $p(\underline{z}|x)\in\mathcal{P}$.

Let $\underline{\hat{X}}\triangleq(\hat{X}_{\mathcal{S}_{k,i}})_{k=1,\cdots,L; i=1,\cdots,2^{k-1}}$ be jointly distributed with $X$ and $\underline{Z}$ such that $p(\underline{\hat{x}}|x)\in\mathcal{P}(\underline{\mathbf{D}})$ and $\underline{Z}\leftrightarrow X\leftrightarrow\underline{\hat{X}}$ form a Markov chain.
Note that
\begin{align}
 I(X; \hat{X}_{\mathcal{S}_{1,1}}|Z_{1,1}) &= I(X; \hat{X}_{\mathcal{S}_{1,1}})-I(Z_{1,1}; \hat{X}_{\mathcal{S}_{1,1}}) \nonumber\\
&=\frac{1}{2}\log \frac{|\mathbf{\Sigma}_X|}{|\mathbf{\Sigma}_X+\mathbf{\Sigma}_{N_{1,1}}|}+I(N_{1,1};X+N_{1,1}|\hat{X}_{\mathcal{S}_{1,1}})\nonumber\\
&\geq  \frac{1}{2} \log  \frac{| \mathbf{\Sigma}_{X}||\mathbf{D}_{\mathcal{S}_{1,1}}+ \mathbf{\Sigma}_{N_{1,1}}|}{ | \mathbf{D}_{\mathcal{S}_{1,1}}||\mathbf{\Sigma}_{X}+ \mathbf{\Sigma}_{N_{1,1}}| }  , \label{eqn:QG1}
 \end{align}
 where (\ref{eqn:QG1}) follows from the conditional version of the worst additive noise lemma \cite{Ihara78,DC01}.
 Similarly, we have
 \begin{align}
 I(Z_{k,i}; \hat{X}_{\mathcal{S}_{k+1,2i-1}} | Z_{k+1,2i-1}) &\geq   \frac{1}{2} \log   \frac{ |\mathbf{\Sigma}_{X}+ \mathbf{\Sigma}_{N_{k,i}}||\mathbf{D}_{\mathcal{S}_{k+1,2i-1}}+ \mathbf{\Sigma}_{N_{k+1,2i-1}}|}{|\mathbf{\Sigma}_{X}+ \mathbf{\Sigma}_{N_{k+1,2i-1}}||\mathbf{D}_{\mathcal{S}_{k+1,2i-1}}+ \mathbf{\Sigma}_{N_{k,i}}|}  ,\nonumber\\
 &\hspace{1in}k=1,\cdots,L-2; i=1,\cdots,2^{k-1}, \label{eqn:QG2} \\
 I(Z_{k,i}; \hat{X}_{\mathcal{S}_{k+1,2i}} | Z_{k+1,2i}) &\geq  \frac{1}{2} \log   \frac{|\mathbf{\Sigma}_{X}+ \mathbf{\Sigma}_{N_{k,i}}||\mathbf{D}_{\mathcal{S}_{k+1,2i}}+ \mathbf{\Sigma}_{N_{k+1,2i}}|}{|\mathbf{\Sigma}_{X}+ \mathbf{\Sigma}_{N_{k+1,2i}}||\mathbf{D}_{\mathcal{S}_{k+1,2i}}+ \mathbf{\Sigma}_{N_{k,i}}|}  ,\nonumber\\
 &\hspace{1in}k=1,\cdots,L-2; i=1,\cdots,2^{k-1}. \label{eqn:QG3}
\end{align}
Moreover, it is clear that
\begin{align}
&I(Z_{L-1,i}; \hat{X}_{\mathcal{S}_{L ,2i-1}}) \geq \frac{1}{2} \log      \frac{|\mathbf{\Sigma}_{X} +\mathbf{\Sigma}_{N_{L-1,i}} |}{|\mathbf{D}_{\mathcal{S}_{L,2i-1}}+\mathbf{\Sigma}_{N_{L-1,i}}|},\quad i=1,\cdots,2^{L-2}, \label{eqn:QG4}\\
&I(Z_{L-1,i}; \hat{X}_{\mathcal{S}_{L ,2i}}) \geq \frac{1}{2} \log      \frac{|\mathbf{\Sigma}_{X} +\mathbf{\Sigma}_{N_{L-1,i}}|}{|\mathbf{D}_{\mathcal{S}_{L,2i}}+\mathbf{\Sigma}_{N_{L-1,i}}|},\quad i=1,\cdots,2^{L-2}. \label{eqn:QG5}
\end{align}
Combining \eqref{eqn:QG1}--\eqref{eqn:QG5} proves
\begin{align}
r_G(\underline{\mathbf{D}})& \geq  \sup\limits_{(\mathbf{\Sigma}_{N_{k,i}})_{k=1,\cdots,L-1; i=1,\cdots,2^{k-1}}}  \frac{1}{2} \log  \frac{| \mathbf{\Sigma}_{X}||\mathbf{D}_{\mathcal{S}_{1,1}}+ \mathbf{\Sigma}_{N_{1,1}}|}{ | \mathbf{D}_{\mathcal{S}_{1,1}}||\mathbf{\Sigma}_{X}+ \mathbf{\Sigma}_{N_{1,1}}| } \nonumber \\
&\hspace{1in}+\sum\limits_{k=1}^{L-2}\sum\limits_{i=1}^{2^{k-1}}\left(\frac{1}{2} \log   \frac{ |\mathbf{\Sigma}_{X}+ \mathbf{\Sigma}_{N_{k,i}}||\mathbf{D}_{\mathcal{S}_{k+1,2i-1}}+ \mathbf{\Sigma}_{N_{k+1,2i-1}}|}{|\mathbf{\Sigma}_{X}+ \mathbf{\Sigma}_{N_{k+1,2i-1}}||\mathbf{D}_{\mathcal{S}_{k+1,2i-1}}+ \mathbf{\Sigma}_{N_{k,i}}|}\right.\nonumber\\
&\hspace{1.7in}\left.+\frac{1}{2} \log   \frac{|\mathbf{\Sigma}_{X}+ \mathbf{\Sigma}_{N_{k,i}}||\mathbf{D}_{\mathcal{S}_{k+1,2i}}+ \mathbf{\Sigma}_{N_{k+1,2i}}|}{|\mathbf{\Sigma}_{X}+ \mathbf{\Sigma}_{N_{k+1,2i}}||\mathbf{D}_{\mathcal{S}_{k+1,2i}}+ \mathbf{\Sigma}_{N_{k,i}}|}\right)\nonumber\\
&\hspace{1in}+\sum\limits_{i=1}^{2^{L-2}}\left(\frac{1}{2} \log      \frac{|\mathbf{\Sigma}_{X} +\mathbf{\Sigma}_{N_{L-1,i}} |}{|\mathbf{D}_{\mathcal{S}_{L,2i-1}}+\mathbf{\Sigma}_{N_{L-1,i}}|}+\frac{1}{2} \log      \frac{|\mathbf{\Sigma}_{X} +\mathbf{\Sigma}_{N_{L-1,i}} |}{|\mathbf{D}_{\mathcal{S}_{L,2i}}+\mathbf{\Sigma}_{N_{L-1,i}}|}\right)\nonumber\\
& \text{subject to}  \quad \mathbf{\Sigma}_{N_{k+1, 2i-1}}, \mathbf{\Sigma}_{N_{k+1, 2i}} \succ \mathbf{\Sigma}_{N_{k,i}} \succ \mathbf{0}, \quad k=1,\cdots,L-2; i=1,\cdots,2^{k-1}. \label{eq:ELB}
\end{align}
It can be verified that
\begin{align*}
&\frac{1}{2} \log  \frac{| \mathbf{\Sigma}_{X}||\mathbf{D}_{\mathcal{S}_{1,1}}+ \mathbf{\Sigma}_{N_{1,1}}|}{ | \mathbf{D}_{\mathcal{S}_{1,1}}||\mathbf{\Sigma}_{X}+ \mathbf{\Sigma}_{N_{1,1}}| } +\sum\limits_{k=1}^{L-2}\sum\limits_{i=1}^{2^{k-1}}\left(\frac{1}{2} \log   \frac{|\mathbf{\Sigma}_{X}+ \mathbf{\Sigma}_{N_{k,i}}||\mathbf{D}_{\mathcal{S}_{k+1,2i-1}}+ \mathbf{\Sigma}_{N_{k+1,2i-1}}|}{|\mathbf{\Sigma}_{X}+ \mathbf{\Sigma}_{N_{k+1,2i-1}}||\mathbf{D}_{\mathcal{S}_{k+1,2i-1}}+ \mathbf{\Sigma}_{N_{k,i}}|}\right.\nonumber\\
&\hspace{2.3in}\left.+\frac{1}{2} \log   \frac{|\mathbf{\Sigma}_{X}+ \mathbf{\Sigma}_{N_{k,i}}||\mathbf{D}_{\mathcal{S}_{k+1,2i}}+ \mathbf{\Sigma}_{N_{k+1,2i}}|}{|\mathbf{\Sigma}_{X}+ \mathbf{\Sigma}_{N_{k+1,2i}}||\mathbf{D}_{\mathcal{S}_{k+1,2i}}+ \mathbf{\Sigma}_{N_{k,i}}|}\right)\nonumber\\
&+\sum\limits_{i=1}^{2^{L-2}}\left(\frac{1}{2} \log      \frac{|\mathbf{\Sigma}_{X} +\mathbf{\Sigma}_{N_{L-1,i}} |}{|\mathbf{D}_{\mathcal{S}_{L,2i-1}}+\mathbf{\Sigma}_{N_{L-1,i}}|}+\frac{1}{2} \log      \frac{|\mathbf{\Sigma}_{X} +\mathbf{\Sigma}_{N_{L-1,i}} |}{|\mathbf{D}_{\mathcal{S}_{L,2i}}+\mathbf{\Sigma}_{N_{L-1,i}}|}\right)\nonumber\\
&=\frac{1}{2} \log  \frac{| \mathbf{\Sigma}_{X}||\mathbf{D}_{\mathcal{S}_{1,1}}+ \mathbf{\Sigma}_{N_{1,1}}|}{ | \mathbf{D}_{\mathcal{S}_{1,1}}||\mathbf{\Sigma}_{X}+ \mathbf{\Sigma}_{N_{1,1}}| }+\sum\limits_{k=1}^{L-1}\sum\limits_{i=1}^{2^{k-1}}\left(\frac{1}{2}\log\frac{|\mathbf{\Sigma}_X+\mathbf{\Sigma}_{N_{k,i}}|}{|\mathbf{D}_{\mathcal{S}_{k+1,2i-1}}+\mathbf{\Sigma}_{N_{k,i}}|}+\frac{1}{2}\log\frac{|\mathbf{\Sigma}_X+\mathbf{\Sigma}_{N_{k,i}}|}{|\mathbf{D}_{\mathcal{S}_{k+1,2i}}+\mathbf{\Sigma}_{N_{k,i}}|}\right)\\
&\quad-\sum\limits_{k=1}^{L-2}\sum\limits_{i=1}^{2^{k-1}}\left(\frac{1}{2}\log\frac{|\mathbf{\Sigma}_{X}+\mathbf{\Sigma}_{N_{k+1,2i-1}}|}{|\mathbf{D}_{\mathcal{S}_{k+1,2i-1}}+\mathbf{\Sigma}_{N_{k+1,2i-1}}|}+\frac{1}{2}\log\frac{|\mathbf{\Sigma}_{X}+\mathbf{\Sigma}_{N_{k+1,2i}}|}{|\mathbf{D}_{\mathcal{S}_{k+1,2i}}+\mathbf{\Sigma}_{N_{k+1,2i}}|}\right)\\
&=\frac{1}{2} \log  \frac{| \mathbf{\Sigma}_{X}|}{ | \mathbf{D}_{\mathcal{S}_{1,1}}| }+\sum\limits_{k=1}^{L-1}\sum\limits_{i=1}^{2^{k-1}}\left(\frac{1}{2}\log\frac{|\mathbf{\Sigma}_X+\mathbf{\Sigma}_{N_{k,i}}|}{|\mathbf{D}_{\mathcal{S}_{k+1,2i-1}}+\mathbf{\Sigma}_{N_{k,i}}|}+\frac{1}{2}\log\frac{|\mathbf{\Sigma}_X+\mathbf{\Sigma}_{N_{k,i}}|}{|\mathbf{D}_{\mathcal{S}_{k+1,2i}}+\mathbf{\Sigma}_{N_{k,i}}|}\right)\\
&\quad-\sum\limits_{k=1}^{L-1}\sum\limits_{i=1}^{2^{k-1}}\frac{1}{2}\log\frac{|\mathbf{\Sigma}_{X}+\mathbf{\Sigma}_{N_{k,i}}|}{|\mathbf{D}_{\mathcal{S}_{k,i}}+\mathbf{\Sigma}_{N_{k,i}}|}\\
&=\frac{1}{2} \log \frac{|\mathbf{\Sigma}_{X}|}{|\mathbf{D}_{\mathcal{S}_{1,1}}|}  + \sum_{k=1}^{L-1} \sum_{i=1}^{2^{k-1}} \frac{1}{2} \log \frac{ |\mathbf{D}_{\mathcal{S}_{k,i}}+\mathbf{\Sigma}_{N_{k,i}}         |  |\mathbf{\Sigma}_{X}+\mathbf{\Sigma}_{N_{k,i}}         | }{   |\mathbf{D}_{\mathcal{S}_{k+1,2i-1}}+\mathbf{\Sigma}_{N_{k,i}}         |     |\mathbf{D}_{\mathcal{S}_{k+1,2i}}+\mathbf{\Sigma}_{N_{k,i}}         | }.
\end{align*}
Now set
\begin{align*}
\mathbf{\Theta}_{k,i} = ( \mathbf{\Sigma}_{X}^{-1}+\mathbf{\Sigma}^{-1}_{N_{k,i}})^{-1}, \quad k=1,\cdots,L-1; i=1,\cdots,2^{k-1}.
\end{align*}
Note that
\begin{align*}
&\mathbf{\Sigma}_{N_{k+1, 2i-1}}, \mathbf{\Sigma}_{N_{k+1, 2i}} \succ \mathbf{\Sigma}_{N_{k,i}} \succ \mathbf{0}\Leftrightarrow \mathbf{\Sigma}_{X} \succ \mathbf{\Theta}_{k+1, 2i-1}, \mathbf{\Theta}_{k+1, 2i} \succ \mathbf{\Theta}_{k,i} \succ \mathbf{0},\\
&\hspace{2.5in} k=1,\cdots,L-2; i=1,\cdots,2^{k-1},
\end{align*}
and
\begin{align*}
&\frac{1}{2} \log \frac{ |\mathbf{D}_{\mathcal{S}_{k,i}}+\mathbf{\Sigma}_{N_{k,i}}         |  |\mathbf{\Sigma}_{X}+\mathbf{\Sigma}_{N_{k,i}}         | }{   |\mathbf{D}_{\mathcal{S}_{k+1,2i-1}}+\mathbf{\Sigma}_{N_{k,i}}         |     |\mathbf{D}_{\mathcal{S}_{k+1,2i}}+\mathbf{\Sigma}_{N_{k,i}}         | } \\
&=\frac{1}{2} \log \frac{ |\mathbf{D}_{\mathcal{S}_{k,i}}+(\mathbf{\Theta}^{-1}_{k,i}- \mathbf{\Sigma}_{X}^{-1})^{-1}         |  |\mathbf{\Sigma}_{X}+(\mathbf{\Theta}^{-1}_{k,i}- \mathbf{\Sigma}_{X}^{-1})^{-1}         | }{   |\mathbf{D}_{\mathcal{S}_{k+1,2i-1}}+(\mathbf{\Theta}^{-1}_{k,i}- \mathbf{\Sigma}_{X}^{-1})^{-1}         |     |\mathbf{D}_{\mathcal{S}_{k+1,2i}}+(\mathbf{\Theta}^{-1}_{k,i}- \mathbf{\Sigma}_{X}^{-1})^{-1}          | } \\
&=\frac{1}{2} \log \frac{ |\mathbf{D}_{\mathcal{S}_{k,i}}(\mathbf{\Theta}^{-1}_{k,i}- \mathbf{\Sigma}_{X}^{-1})+\mathbf{I}_m         |  |\mathbf{\Sigma}_{X}(\mathbf{\Theta}^{-1}_{k,i}- \mathbf{\Sigma}_{X}^{-1})+\mathbf{I}_m     | }{   |\mathbf{D}_{\mathcal{S}_{k+1,2i-1}}(\mathbf{\Theta}^{-1}_{k,i}- \mathbf{\Sigma}_{X}^{-1})+\mathbf{I}_m|     |\mathbf{D}_{\mathcal{S}_{k+1,2i}}(\mathbf{\Theta}^{-1}_{k,i}- \mathbf{\Sigma}_{X}^{-1})+\mathbf{I}_m | }\\
&=\frac{1}{2} \log \frac{  |\mathbf{\Sigma}_{X}| |\mathbf{D}_{\mathcal{S}_{k,i}}\mathbf{\Sigma}_{X}^{-1}(\mathbf{\Sigma}_{X} - \mathbf{\Theta}_{k,i} ) +  \mathbf{\Theta}_{k,i} |      }{ |\mathbf{D}_{\mathcal{S}_{k+1,2i-1}}\mathbf{\Sigma}_{X}^{-1}(\mathbf{\Sigma}_{X} - \mathbf{\Theta}_{k,i} ) +  \mathbf{\Theta}_{k,i} |   |\mathbf{D}_{\mathcal{S}_{k+1,2i}}\mathbf{\Sigma}_{X}^{-1}(\mathbf{\Sigma}_{X} - \mathbf{\Theta}_{k,i} ) +  \mathbf{\Theta}_{k,i} |       } ,\\
  &\hspace{2.5in}k=1,\cdots,L-1; i=1,\cdots,2^{k-1},
\end{align*}
where $\mathbf{I}_m$ denotes the $m\times m$ identity matrix.
Therefore, we can write \eqref{eq:ELB} equivalently as
\begin{align}
r_G(\underline{\mathbf{D}})&\geq \sup\limits_{\underline{\mathbf{\Theta}}} \frac{1}{2} \log \frac{|\mathbf{\Sigma}_{X}|}{|\mathbf{D}_{\mathcal{S}_{1,1}}|}  \nonumber \\
& \qquad + \sum_{k=1}^{L-1} \sum_{i=1}^{2^{k-1}} \frac{1}{2} \log \frac{  |\mathbf{\Sigma}_{X}| |\mathbf{D}_{\mathcal{S}_{k,i}}\mathbf{\Sigma}_{X}^{-1}(\mathbf{\Sigma}_{X} - \mathbf{\Theta}_{k,i}   ) +  \mathbf{\Theta}_{k,i} |      }{ |\mathbf{D}_{\mathcal{S}_{k+1,2i-1}}\mathbf{\Sigma}_{X}^{-1}(\mathbf{\Sigma}_{X} - \mathbf{\Theta}_{k,i}  ) +  \mathbf{\Theta}_{k,i} |   |\mathbf{D}_{\mathcal{S}_{k+1,2i}}\mathbf{\Sigma}_{X}^{-1}(\mathbf{\Sigma}_{X} - \mathbf{\Theta}_{k,i}  ) +  \mathbf{\Theta}_{k,i} |       }  \nonumber \\
& \text{subject to} \quad\mathbf{\Sigma}_{X} \succ \mathbf{\Theta}_{k+1, 2i-1}, \mathbf{\Theta}_{k+1, 2i} \succ \mathbf{\Theta}_{k,i} \succ \mathbf{0}, \quad k=1,\cdots,L-2; i=1,\cdots,2^{k-1},\nonumber
\end{align}
from which the desired result follows immediately.

\subsection{The Achievability Part}\label{sec:achievability}

First consider the case where $\mathbf{0}\prec\mathbf{D}_{\mathcal{S}_{k,i}} \prec \mathbf{\Sigma}_{X}$, $k=1,\cdots,L; i=1,\cdots,2^{k-1}$. In this case, we have
\begin{align*}
\mathbf{\Sigma}_{\mathcal{S}_{k,i}}\triangleq(\mathbf{D}^{-1}_{\mathcal{S}_{k,i}}-\mathbf{\Sigma}^{-1}_{X})^{-1}\succ\mathbf{0},\quad k=1,\cdots,L; i=1,\cdots,2^{k-1}.
\end{align*}
It can be verified that
\begin{align}
 \frac{1}{2} \log \frac{|\mathbf{\Sigma_{X}}|}{|\mathbf{D}_{\mathcal{S}_{1,1}}|}&=\frac{1}{2} \log |\mathbf{\Sigma_{X}}( \mathbf{\Sigma}_{\mathcal{S}_{1,1}}^{-1} + \mathbf{\Sigma}_{X}^{-1} )| \nonumber \\
&= \frac{1}{2} \log \frac{|\mathbf{\Sigma_{X}} + \mathbf{\Sigma}_{\mathcal{S}_{1,1}}|}{|\mathbf{\Sigma}_{\mathcal{S}_{1,1}}|}. \label{eq:tran1}
\end{align}
Moreover,
\begin{align}
 & \frac{1}{2} \log \frac{|\mathbf{\Sigma}_{X}|}{ |\mathbf{D}_{\mathcal{S}_{k,i}}\mathbf{\Sigma}_{X}^{-1}(\mathbf{\Sigma}_{X} - \mathbf{\Theta}_{k,i} ) +  \mathbf{\Theta}_{k,i}|} \nonumber \\
&= \frac{1}{2} \log \frac{|\mathbf{\Sigma}_{X}|}{ |( \mathbf{\Sigma}_{\mathcal{S}_{k,i}}^{-1} + \mathbf{\Sigma}_{X}^{-1}  )^{-1}\mathbf{\Sigma}_{X}^{-1}(\mathbf{\Sigma}_{X} - \mathbf{\Theta}_{k,i} ) +  \mathbf{\Theta}_{k,i}|} \nonumber \\
&= \frac{1}{2} \log \frac{|\mathbf{\Sigma}_{X}|}{|   \mathbf{\Sigma}_{\mathcal{S}_{k,i}} ( \mathbf{\Sigma}_{X} + \mathbf{\Sigma}_{\mathcal{S}_{k,i}})^{-1} (\mathbf{\Sigma}_{X} - \mathbf{\Theta}_{k,i} ) +  \mathbf{\Theta}_{k,i}                |} \nonumber \\
&= \frac{1}{2} \log \frac{|\mathbf{\Sigma}_{X}|}{|   \mathbf{\Sigma}_{\mathcal{S}_{k,i}} ( \mathbf{\Sigma}_{X} + \mathbf{\Sigma}_{\mathcal{S}_{k,i}})^{-1} (\mathbf{\Sigma}_{X} - \mathbf{\Theta}_{k,i} ) - ( \mathbf{\Sigma}_{X}-\mathbf{\Theta}_{k,i}) +\mathbf{\Sigma}_{X}     |} \nonumber \\
&= \frac{1}{2} \log \frac{|\mathbf{\Sigma}_{X}|}{| \mathbf{\Sigma}_{X} - \mathbf{\Sigma}_{X}( \mathbf{\Sigma}_{X} + \mathbf{\Sigma}_{\mathcal{S}_{k,i}})^{-1} (\mathbf{\Sigma}_{X} - \mathbf{\Theta}_{k,i}  )|} \nonumber \\
&= \frac{1}{2} \log \frac{|\mathbf{\Sigma}_{X} + \mathbf{\Sigma}_{\mathcal{S}_{k,i}}|}{| \mathbf{\Sigma}_{X} + \mathbf{\Sigma}_{\mathcal{S}_{k,i}}- \mathbf{\Sigma}_{X} + \mathbf{\Theta}_{k,i}   |} \nonumber \\
&= \frac{1}{2} \log \frac{|\mathbf{\Sigma}_{X} + \mathbf{\Sigma}_{\mathcal{S}_{k,i}}|}{|  \mathbf{\Theta}_{k,i} + \mathbf{\Sigma}_{\mathcal{S}_{k,i}} |}, \quad k=1,\cdots,L-1; i=1,\cdots,2^{k-1}, \label{eq:tran2}
\end{align}
and, similarly,
\begin{align}
& \frac{1}{2} \log \frac{|\mathbf{\Sigma}_{X}|}{ \left|\mathbf{D}_{\mathcal{S}_{k+1,2i-1}}\mathbf{\Sigma}_{X}^{-1}(\mathbf{\Sigma}_{X} - \mathbf{\Theta}_{k,i} ) +  \mathbf{\Theta}_{k,i}\right|} \nonumber \\
&= \frac{1}{2} \log \frac{|\mathbf{\Sigma}_{X} + \mathbf{\Sigma}_{\mathcal{S}_{k+1,2i-1}}|}{|  \mathbf{\Theta}_{k,i} + \mathbf{\Sigma}_{\mathcal{S}_{k+1,2i-1}} |}, \quad k=1,\cdots,L-1; i=1,\cdots,2^{k-1}, \label{eq:tran3}\\
& \frac{1}{2} \log \frac{|\mathbf{\Sigma}_{X}|}{ |\mathbf{D}_{\mathcal{S}_{k+1,2i}}\mathbf{\Sigma}_{X}^{-1}(\mathbf{\Sigma}_{X} - \mathbf{\Theta}_{k,i}  ) +  \mathbf{\Theta}_{k,i}|} \nonumber \\
&= \frac{1}{2} \log \frac{|\mathbf{\Sigma}_{X} + \mathbf{\Sigma}_{\mathcal{S}_{k+1,2i}}|}{|  \mathbf{\Theta}_{k,i} + \mathbf{\Sigma}_{\mathcal{S}_{k+1,2i}} |}, \quad k=1,\cdots,L-1; i=1,\cdots,2^{k-1}. \label{eq:tran4}
\end{align}
In view of (\ref{eq:tran1})-(\ref{eq:tran4}), the maximization problem in (\ref{eq:main}) can be expressed equivalently as
\begin{align}
& \max\limits_{\underline{\mathbf{\Theta}}} \frac{1}{2} \log\frac{|\mathbf{\Sigma_{X}} + \mathbf{\Sigma}_{\mathcal{S}_{1,1}}|}{|\mathbf{\Sigma}_{\mathcal{S}_{1,1}}|}   + \sum_{k=1}^{L-1} \sum_{i=1}^{2^{k-1}} \frac{1}{2} \log\frac{|  \mathbf{\Theta}_{k,i} + \mathbf{\Sigma}_{\mathcal{S}_{k,i}} ||\mathbf{\Sigma}_{X} + \mathbf{\Sigma}_{\mathcal{S}_{k+1,2i-1}}||\mathbf{\Sigma}_{X} + \mathbf{\Sigma}_{\mathcal{S}_{k+1,2i}}|}{|\mathbf{\Sigma}_{X} + \mathbf{\Sigma}_{\mathcal{S}_{k,i}}||  \mathbf{\Theta}_{k,i} + \mathbf{\Sigma}_{\mathcal{S}_{k+1,2i-1}} ||  \mathbf{\Theta}_{k,i} + \mathbf{\Sigma}_{\mathcal{S}_{k+1,2i}}|}  \nonumber \\
& \text{subject to} \quad\mathbf{\Sigma}_{X} \succeq \mathbf{\Theta}_{k+1, 2i-1}, \mathbf{\Theta}_{k+1, 2i} \succeq \mathbf{\Theta}_{k,i} \succeq \mathbf{0}, \quad k=1,\cdots,L-2; i=1,\cdots,2^{k-1}.\label{eq:mainequiv}
\end{align}
Let $\underline{\mathbf{\Theta}}^*\triangleq(\mathbf{\Theta}^*_{k,i})_{k=1,\cdots,L-1; i=1,\cdots,2^{k-1}}$ be an optimal solution of the above maximization problem. The following lemma provides the Karush-Kuhn-Tucker conditions that $\underline{\mathbf{\Theta}}^*$ needs to satisfy. The proof can be found in Appendix \ref{app:KKT}.
\begin{lemma}\label{lemma:KKT}
There exist $\mathbf{M}^*_{k,i}$, $k=1,\cdots,L; i=1,\cdots,2^{k-1}$, such that
\begin{align}
&( \mathbf{\Theta}_{k,i}^{*} +\mathbf{\Sigma}_{\mathcal{S}_{k+1,2i-1}})^{-1} + \mathbf{M}^*_{k+1,2i-1} + ( \mathbf{\Theta}_{k,i}^{*} +\mathbf{\Sigma}_{\mathcal{S}_{k+1,2i}})^{-1} + \mathbf{M}^*_{k+1,2i} = ( \mathbf{\Theta}_{k,i}^{*} +\mathbf{\Sigma}_{\mathcal{S}_{k,i}})^{-1} + \mathbf{M}^*_{k,i}, \nonumber \\
&\hspace{3.5in}     k=1,\cdots,L-1; i=1,\cdots,2^{k-1}, \label{eq:KKT1}\\
& \mathbf{M}^*_{1,1}\mathbf{\Theta}_{1,1}^{*} = \mathbf{\Theta}_{1,1}^{*}\mathbf{M}^*_{1,1}  = \mathbf{0}, \label{eq:KKT2}\\
& \mathbf{M}^*_{k+1,2i-1} (\mathbf{\Theta}_{k+1,2i-1}^{*} - \mathbf{\Theta}_{k,i}^{*}) =  (\mathbf{\Theta}_{k+1,2i-1}^{*} - \mathbf{\Theta}_{k,i}^{*}) \mathbf{M}^*_{k+1,2i-1}  = \mathbf{0}, \nonumber\\
&\hspace{2in} k=1,\cdots,L-2; i=1,\cdots,2^{k-1}, \label{eq:KKT3} \\
& \mathbf{M}^*_{k+1,2i} (\mathbf{\Theta}_{k+1,2i}^{*} - \mathbf{\Theta}_{k,i}^{*}) =  (\mathbf{\Theta}_{k+1,2i}^{*} - \mathbf{\Theta}_{k,i}^{*}) \mathbf{M}^*_{k+1,2i}  = \mathbf{0}, \nonumber\\
&\hspace{2in} k=1,\cdots,L-2; i=1,\cdots,2^{k-1}, \nonumber \\
& \mathbf{M}^*_{L,2i-1} (\mathbf{\Sigma}_{X} - \mathbf{\Theta}_{L-1,i}^{*}) =  (\mathbf{\Sigma}_{X} - \mathbf{\Theta}_{L-1,i}^{*}) \mathbf{M}^*_{L,2i-1}  = \mathbf{0}, \quad i=1,\cdots,2^{L-2}, \label{eq:KKT4}\\
&\mathbf{M}^*_{k,i}  \succeq \mathbf{0}, \quad k=1, \cdots, L-1; i=1, \cdots, 2^{k-1},\nonumber \\
& \mathbf{M}^*_{L,2i-1} \succeq \mathbf{0}, \quad i=1,\cdots,2^{L-2}, \nonumber\\
& \mathbf{M}^*_{L,2i} = \mathbf{0}, \quad i=1,\cdots,2^{L-2}.\nonumber
\end{align}
\end{lemma}



Inspired by the enhancement argument in \cite{WSS06} (see also \cite{WV07,WV09}), we define $\underline{\tilde{\mathbf{\Sigma}}}=(\tilde{\mathbf{\Sigma}}_{\mathcal{S}_{k,i}})_{k=1,\cdots,L-1;i=1,\cdots,2^{k-1}}$ with
\begin{align}
&\tilde{\mathbf{\Sigma}}_{\mathcal{S}_{k,i}}=(( \mathbf{\Theta}_{k,i}^{*} +\mathbf{\Sigma}_{\mathcal{S}_{k,i}})^{-1} + \mathbf{M}^*_{k,i})^{-1}- \mathbf{\Theta}_{k,i}^{*}, \quad
     k=1,\cdots,L-1; i=1,\cdots,2^{k-1}, \label{eq:en1} \\
&\tilde{\mathbf{\Sigma}}_{\mathcal{S}_{L,i}}=(( \mathbf{\Sigma}_{X} +{\mathbf{\Sigma}}_{\mathcal{S}_{L,i}})^{-1} + \mathbf{M}^*_{L,i})^{-1}-\mathbf{\Sigma}_{X}, \quad i=1,\cdots,2^{L-1}. \label{eq:en2}
\end{align}
The next lemma collects some useful facts about $\underline{\tilde{\mathbf{\Sigma}}}$. The proof is relegated to Appendix \ref{enhance}.

\begin{lemma}\label{lemma:enhance}
$\underline{\tilde{\mathbf{\Sigma}}}$ has the following properties:
\begin{align}
&( \mathbf{\Theta}_{k,i}^{*} +\tilde{\mathbf{\Sigma}}_{\mathcal{S}_{k+1,2i-1}})^{-1} = ( \mathbf{\Theta}_{k,i}^{*} +\mathbf{\Sigma}_{\mathcal{S}_{k+1,2i-1}})^{-1}+\mathbf{M}^*_{k+1,2i-1}, \quad k=1,\cdots,L-1; i=1,\cdots,2^{k-1}, \label{eq:enhance1}\\
&( \mathbf{\Theta}_{k,i}^{*} +\tilde{\mathbf{\Sigma}}_{\mathcal{S}_{k+1,2i}})^{-1} = ( \mathbf{\Theta}_{k,i}^{*} +\mathbf{\Sigma}_{\mathcal{S}_{k+1,2i}})^{-1}+\mathbf{M}^*_{k+1,2i}, \quad k=1,\cdots,L-1; i=1,\cdots,2^{k-1}, \label{eq:enhance2}\\
&( \mathbf{\Theta}_{k,i}^{*} +\tilde{\mathbf{\Sigma}}_{\mathcal{S}_{k+1,2i-1}})^{-1}  + ( \mathbf{\Theta}_{k,i}^{*} +\tilde{\mathbf{\Sigma}}_{\mathcal{S}_{k+1,2i}})^{-1}  = ( \mathbf{\Theta}_{k,i}^{*} +\tilde{\mathbf{\Sigma}}_{\mathcal{S}_{k,i}})^{-1}, \nonumber \\
 &\hspace{3.5in}    k=1,\cdots,L-1; i=1,\cdots,2^{k-1}, \label{eq:KKT_en1}\\
&\tilde{\mathbf{\Sigma}}_{\mathcal{S}_{k+1,2i-1}},\tilde{\mathbf{\Sigma}}_{\mathcal{S}_{k+1,2i}} \succ  \tilde{\mathbf{\Sigma}}_{\mathcal{S}_{k,i}} \succ \mathbf{0}, \quad  k=1,\cdots,L-1; i=1,\cdots,2^{k-1}, \label{eq:enhance8}\\
&\tilde{\mathbf{\Sigma}}_{\mathcal{S}_{1,1}}^{-1} ( \mathbf{\Theta}_{1,1}^{*} + \tilde{\mathbf{\Sigma}}_{\mathcal{S}_{1,1}})=\mathbf{\Sigma}_{\mathcal{S}_{1,1}}^{-1} ( \mathbf{\Theta}_{1,1}^{*} + \mathbf{\Sigma}_{\mathcal{S}_{1,1}}), \label{eq:enhance3}\\
&(\mathbf{\Theta}_{k,i}^{*}+ \tilde{\mathbf{\Sigma}}_{\mathcal{S}_{k+1,2i-1}} )^{-1}( \mathbf{\Theta}_{k+1,2i-1}^{*}+ \tilde{\mathbf{\Sigma}}_{\mathcal{S}_{k+1,2i-1}} ) = (\mathbf{\Theta}_{k,i}^{*}+ \mathbf{\Sigma}_{\mathcal{S}_{k+1,2i-1}} )^{-1}( \mathbf{\Theta}_{k+1,2i-1}^{*}+ \mathbf{\Sigma}_{\mathcal{S}_{k+1,2i-1}} ) \nonumber \\
&\hspace{3.5in} k=1,\cdots,L-2; i=1,\cdots,2^{k-1}, \label{eq:enhance4}\\
& (\mathbf{\Theta}_{k,i}^{*}+ \tilde{\mathbf{\Sigma}}_{\mathcal{S}_{k+1,2i}} )^{-1}( \mathbf{\Theta}_{k+1,2i}^{*}+ \tilde{\mathbf{\Sigma}}_{\mathcal{S}_{k+1,2i}} ) = (\mathbf{\Theta}_{k,i}^{*}+ \mathbf{\Sigma}_{\mathcal{S}_{k+1,2i}})^{-1}( \mathbf{\Theta}_{k+1,2i}^{*}+ \mathbf{\Sigma}_{\mathcal{S}_{k+1,2i}} ) \nonumber \\
&\hspace{3.5in} k=1,\cdots,L-2; i=1,\cdots,2^{k-1}, \label{eq:enhance5}\\
&(\mathbf{\Theta}_{L-1,i}^{*}+\tilde{\mathbf{\Sigma}}_{\mathcal{S}_{L,2i-1}})^{-1} (\mathbf{\Sigma}_{X}+\tilde{\mathbf{\Sigma}}_{\mathcal{S}_{L,2i-1}}) =  (\mathbf{\Theta}_{L-1,i}^{*}+{\mathbf{\Sigma}}_{\mathcal{S}_{L,2i-1}})^{-1} (\mathbf{\Sigma}_{X}+{\mathbf{\Sigma}}_{\mathcal{S}_{L,2i-1}}), \nonumber \\
&\hspace{3.5in} i=1,\cdots,2^{L-2}. \label{eq:enhance6}\\
& (\mathbf{\Theta}_{L-1,i}^{*}+\tilde{\mathbf{\Sigma}}_{\mathcal{S}_{L,2i}})^{-1} (\mathbf{\Sigma}_{X}+\tilde{\mathbf{\Sigma}}_{\mathcal{S}_{L,2i}}) =  (\mathbf{\Theta}_{L-1,i}^{*}+{\mathbf{\Sigma}}_{\mathcal{S}_{L,2i}})^{-1} (\mathbf{\Sigma}_{X}+{\mathbf{\Sigma}}_{\mathcal{S}_{L,2i}}), \nonumber \\
&\hspace{3.5in} i=1,\cdots,2^{L-2}. \label{eq:enhance7}
\end{align}
\end{lemma}




\begin{lemma}\label{lem:sumrate}
We have
\begin{align*}
R^*_G(\underline{\mathbf{D}})=\frac{1}{2} \log\frac{|\mathbf{\Sigma_{X}} + \tilde{\mathbf{\Sigma}}_{\mathcal{S}_{1,1}}|}{|\tilde{\mathbf{\Sigma}}_{\mathcal{S}_{1,1}}|}   + \sum_{k=1}^{L-1} \sum_{i=1}^{2^{k-1}} \frac{1}{2} \log\frac{|  \mathbf{\Theta}^*_{k,i} + \tilde{\mathbf{\Sigma}}_{\mathcal{S}_{k,i}} ||\mathbf{\Sigma}_{X} + \tilde{\mathbf{\Sigma}}_{\mathcal{S}_{k+1,2i-1}}||\mathbf{\Sigma}_{X} + \tilde{\mathbf{\Sigma}}_{\mathcal{S}_{k+1,2i}}|}{|\mathbf{\Sigma}_{X} + \tilde{\mathbf{\Sigma}}_{\mathcal{S}_{k,i}}||  \mathbf{\Theta}^*_{k,i} + \tilde{\mathbf{\Sigma}}_{\mathcal{S}_{k+1,2i-1}} ||  \mathbf{\Theta}^*_{k,i} + \tilde{\mathbf{\Sigma}}_{\mathcal{S}_{k+1,2i}}|}.
\end{align*}
\end{lemma}
\begin{IEEEproof}
See Appendix \ref{app:sumrate}.
\end{IEEEproof}

Let $U_j=X+Q_{L,j}$, $j=1,\cdots,M$, where $Q_{L,1},\cdots,Q_{L,M}$ are jointly Gaussian and  independent of $X$. A straightforward extension of the El Gamal-Cover scheme (see, e.g., \cite{C09}) shows\footnote{Roughly speaking, $U_j$ corresponds to the $j$-th description, $j=1,\cdots,M$; moreover,  $\sum_{j=1 }^Mh(U_{j}) -h(U_1,\cdots,U_M | X)$ is the sum rate, and $\cov(X|(U_j)_{j\in\mathcal{S}_{i,k}})$ is the reconstruction covariance distortion based on the subset of descriptions specified by $\mathcal{S}_{k,i}$, $k=1,\cdots,L; i=1,\cdots,2^{k-1}$.} that
\begin{align*}
R_G(\underline{\mathbf{D}})\leq\sum\limits_{j=1 }^Mh(U_{j}) -h(U_1,\cdots,U_M | X)
\end{align*}
if $\cov(X|(U_j)_{j\in\mathcal{S}_{i,k}})\preceq\mathbf{D}_{\mathcal{S}_{k,i}}$, $k=1,\cdots,L; i=1,\cdots,2^{k-1}$.

Now we shall leverage $\underline{\mathbf{\Theta}}^*$ and $\underline{\tilde{\mathbf{\Sigma}}}$ to construct the covariance matrix of $(Q_1,\cdots,Q_M)$ such that the resulting scheme satisfies the distortion constraints and attains the minimum sum rate. Define
\begin{align*}
&\mathbf{\Lambda}_{k,i}=\left(
                \begin{array}{cc}
                  \tilde{\mathbf{\Sigma}}_{\mathcal{S}_{k+1,2i-1}}-\tilde{\mathbf{\Sigma}}_{\mathcal{S}_{k,i}} & -\mathbf{\Theta}^*_{k,i}-\tilde{\mathbf{\Sigma}}_{\mathcal{S}_{k,i}}\\
                  -\mathbf{\Theta}^*_{k,i}-\tilde{\mathbf{\Sigma}}_{\mathcal{S}_{k,i}} & \tilde{\mathbf{\Sigma}}_{\mathcal{S}_{k+1,2i}}-\tilde{\mathbf{\Sigma}}_{\mathcal{S}_{k,i}} \\
                \end{array}
              \right),\\
&\mathbf{\Gamma}_{k,i}=
\left(
                \begin{array}{cc}
\tilde{\mathbf{\Sigma}}_{\mathcal{S}_{k+1,2i-1}} & -\mathbf{\Theta}_{k,i}^{*} \\
  -\mathbf{\Theta}_{k,i}^{*} & \tilde{\mathbf{\Sigma}}_{\mathcal{S}_{k+1,2i}}\\
\end{array}
\right)
\end{align*}
for $k=1,\cdots,L-1;i=1,\cdots,2^{k-1}$.
\begin{lemma}\label{lemma:GTC}
\begin{enumerate}
\item For $k=1,\cdots,L-1;i=1,\cdots,2^{k-1}$,
\begin{align}
\mathbf{\Lambda}_{k,i}\succeq\mathbf{0}.\label{eq:possemdef}
\end{align}

\item For $k=1,\cdots,L-1;i=1,\cdots,2^{k-1}$, there exist $\mathbf{H}_{k+1,2i-1}$ and $\mathbf{H}_{k+1,2i}$ such that
\begin{align}
&\mathbf{H}_{k+1,2i-1}+\mathbf{H}_{k+1,2i}=\mathbf{I}_m,\label{eq:im}\\
&(\mathbf{H}_{k+1,2i-1},\mathbf{H}_{k+1,2i})\mathbf{\Lambda}_{k,i}(\mathbf{H}_{k+1,2i-1},\mathbf{H}_{k+1,2i})^T=\mathbf{0}.\label{eq:zero}
\end{align}
\end{enumerate}
\end{lemma}
\begin{IEEEproof}
See Appendix \ref{app:GTC}.
\end{IEEEproof}

Let $\underline{V}\triangleq(V_{k,i})_{k=1,\cdots,L;i=1,\cdots,2^{k-1}}$ be jointly distributed with $X$ such that
\begin{align*}
p(x,\underline{v})=p(x)p(v_{1,1})\prod\limits_{k=1}^{L-1}\prod\limits_{i=1}^{2^{k-1}}p(v_{k+1,2i-1},v_{k+1,2i}).
\end{align*}
Moreover, we assume that $V_{k,i}$, $k=1,\cdots,L;i=1,\cdots,2^{k-1}$, are $m\times 1$ zero-mean jointly Gaussian random vectors with\footnote{Note that $\tilde{\mathbf{\Sigma}}_{\mathcal{S}_{1,1}}$ and $\mathbf{\Lambda}_{k,i}$, $k=1,\cdots,L-1;i=1,\cdots,2^{k-1}$, are valid covariance matrices as ensured by (\ref{eq:enhance8}) and (\ref{eq:possemdef}).}
\begin{align*}
&\cov(V_{1,1})=\tilde{\mathbf{\Sigma}}_{\mathcal{S}_{1,1}},\\
&\cov(V_{k+1,2i-1},V_{k+1,2i})=\mathbf{\Lambda}_{k,i},\quad k=1,\cdots,L-1;i=1,\cdots,2^{k-1}.
\end{align*}
In this way, the joint distribution of $X$ and $\underline{V}$ is completely specified. Set
\begin{align*}
&Q_{1,1}=V_{1,1},\\
&Q_{k+1,2i-1}=Q_{k,i}+V_{k+1,2i-1},\quad k=1,\cdots,L-1;i=1,\cdots,2^{k-1},\\
&Q_{k+1,2i}=Q_{k,i}+V_{k+1,2i},\quad k=1,\cdots,L-1;i=1,\cdots,2^{k-1}.
\end{align*}
It is easy to verify that
\begin{align}
\cov(Q_{k+1,2i-1},Q_{k+1,2i})=\mathbf{\Gamma}_{k,i},\quad k=1,\cdots,L-1; i=1,\cdots,2^{k-1}.\label{eq:conditionalind}
\end{align}

Note that the constructed $(U_1,\cdots,U_M)$ satisfies the covariance distortion constraints. Indeed,
\begin{align}
\cov(X|(U_j)_{j\in\mathcal{S}_{k,i}})&=\cov(X|X+Q_{k,i})\label{eq:duetopre}\\
&=(\mathbf{\Sigma}^{-1}_{X}+\tilde{\mathbf{\Sigma}}^{-1}_{\mathcal{S}_{k,i}})^{-1}\nonumber\\
&\preceq (\mathbf{\Sigma}^{-1}_{X}+\mathbf{\Sigma}^{-1}_{\mathcal{S}_{k,i}})^{-1}\nonumber\\
&=\mathbf{D}_{\mathcal{S}_{k,i}},\quad k=1,\cdots,L;i=1,\cdots,2^{k-1},\nonumber
\end{align}
where (\ref{eq:duetopre}) is due to (\ref{eq:im}) and (\ref{eq:zero}).
Now it remains to show that the resulting sum rate coincides with $R^*_G(\underline{\mathbf{D}})$. We have
\begin{align}
& \sum\limits_{j =1}^Mh(U_{j}) -h(U_{1},\cdots,U_M |X)  \nonumber\\
&= I(X; (U_{j})_{j \in \mathcal{S}_{1,1}})+ \sum_{k=1}^{{L-1}} \sum_{i=1}^{2^{k-1}} I((U_{j})_{j \in \mathcal{S}_{k+1,2i-1}} ;(U_{j})_{j \in \mathcal{S}_{k+1,2i}}) \label{eq:succentropy}\\
&= I(X; X+Q_{1,1} )+ \sum_{k=1}^{{L-1}} \sum_{i=1}^{2^{k-1}} I(X+Q_{k+1,2i-1} ;X+Q_{k+1,2i}) \label{eq:useQ}\\
&=\frac{1}{2}\log\frac{|\mathbf{\Sigma}_X+\tilde{\mathbf{\Sigma}}_{\mathcal{S}_{1,1}}|}{|\tilde{\mathbf{\Sigma}}_{\mathcal{S}_{1,1}}|}+ \sum_{k=1}^{{L-1}} \sum_{i=1}^{2^{k-1}} I(X+Q_{k+1,2i-1} ;X+Q_{k+1,2i}),\label{eq:sum_rate0}
\end{align}
where (\ref{eq:succentropy}) follows by successively applying the fact that
\begin{align*}
\sum\limits_{i=1}^{2^{k}}h((U_j)_{j \in \mathcal{S}_{k+1, i}} )=\sum\limits_{i=1}^{2^{k-1}}h((U_j)_{j \in \mathcal{S}_{k, i}} )+\sum\limits_{i=1}^{2^{k-1}}I((U_j)_{j \in \mathcal{S}_{k+1, 2i-1}};(U_j)_{j \in \mathcal{S}_{k+1, 2i}} )
\end{align*}
from $k=L-1$ to $k=1$, and (\ref{eq:useQ}) is due to (\ref{eq:im}) and (\ref{eq:zero}). Write $X=\tilde{X}_{{k,i}}+\hat{X}_{k,i}$, where $\tilde{X}_{{k,i}}$ and $\hat{X}_{k,i}$ two $m \times 1$ zero-mean Gaussian random vectors with covariance matrices $\mathbf{\Sigma}_{X} - \mathbf{\Theta}_{k,i}^{*}$ and $\mathbf{\Theta}_{k,i}^{*}$, respectively, $k=1, \cdots, L-1; i=1, \cdots, 2^{k-1}$; moreover, $\tilde{X}_{{k,i}}$, $\hat{X}_{k,i}$, and $(Q_{k+1,2i-1},Q_{k+1,2i})$ are mutually independent, $k=1, \cdots, L-1; i=1, \cdots, 2^{k-1}$. In view of (\ref{eq:conditionalind}),
\begin{align*}
\mathbb{E}[(\hat{X}_{k,i}+Q_{k+1,2i-1})(\hat{X}_{k,i}+Q_{k+1,2i})^T]=\mathbf{0},\quad k=1, \cdots, L-1; i=1, \cdots, 2^{k-1},
\end{align*}
which implies
\begin{align}
I(\hat{X}_{k,i}+Q_{k+1,2i-1};\hat{X}_{k,i}+Q_{k+1,2i})=0,\quad k=1, \cdots, L-1; i=1, \cdots, 2^{k-1}.\label{eq:keyindep}
\end{align}
It can be verified that
\begin{align}
&I(X+Q_{k+1,2i-1} ;X+Q_{k+1,2i})\nonumber\\
&=I(\tilde{X}_{{k,i}};X+Q_{k+1,2i-1})+I(\tilde{X}_{{k,i}};X+Q_{k+1,2i})\nonumber\\
&\quad-I(\tilde{X}_{{k,i}};X+Q_{k+1,2i-1},X+Q_{k+1,2i-1})+I(X+Q_{k+1,2i-1} ;X+Q_{k+1,2i}|\tilde{X}_{{k,i}})\nonumber\\
&=I(\tilde{X}_{{k,i}};X+Q_{k+1,2i-1})+I(\tilde{X}_{{k,i}};X+Q_{k+1,2i})-I(\tilde{X}_{{k,i}};X+Q_{k,i})\nonumber\\
&\quad+I(X+Q_{k+1,2i-1} ;X+Q_{k+1,2i}|\tilde{X}_{{k,i}})\label{eq:uuse1}\\
&=I(\tilde{X}_{{k,i}};X+Q_{k+1,2i-1})+I(\tilde{X}_{{k,i}};X+Q_{k+1,2i})-I(\tilde{X}_{{k,i}};X+Q_{k,i})\nonumber\\
&\quad+I(\hat{X}_{k,i}+Q_{k+1,2i-1};\hat{X}_{k,i}+Q_{k+1,2i})\nonumber\\
&=I(\tilde{X}_{{k,i}};X+Q_{k+1,2i-1})+I(\tilde{X}_{{k,i}};X+Q_{k+1,2i})-I(\tilde{X}_{{k,i}};X+Q_{k,i})\label{eq:uuse2}\\
&=\frac{1}{2}\log\frac{|\mathbf{\Sigma}_X+\tilde{\mathbf{\Sigma}}_{\mathcal{S}_{k+1,2i-1}}|}{|\mathbf{\Theta}^*_{k,i}+\tilde{\mathbf{\Sigma}}_{\mathcal{S}_{k+1,2i-1}}|}+\frac{1}{2}\log\frac{|\mathbf{\Sigma}_X+\tilde{\mathbf{\Sigma}}_{\mathcal{S}_{k+1,2i}}|}{|\mathbf{\Theta}^*_{k,i}+\tilde{\mathbf{\Sigma}}_{\mathcal{S}_{k+1,2i}}|}-\frac{1}{2}\log\frac{|\mathbf{\Sigma}_X+\tilde{\mathbf{\Sigma}}_{\mathcal{S}_{k,i}}|}{|\mathbf{\Theta}^*_{k,i}+\tilde{\mathbf{\Sigma}}_{\mathcal{S}_{k,i}}|}\nonumber\\
&\hspace{2in} k=1,\cdots,L-1; i=1,\cdots,2^{k-1},\label{eq:uuse3}
\end{align}
where (\ref{eq:uuse1}) is due to (\ref{eq:im}) and (\ref{eq:zero}) while (\ref{eq:uuse2}) is due to (\ref{eq:keyindep}). Substituting (\ref{eq:uuse3}) into (\ref{eq:sum_rate0}) and invoking Lemma \ref{lem:sumrate} shows
\begin{align*}
\sum\limits_{j =1}^Mh(U_{j}) -h(U_{1},\cdots,U_M |X)=R^*_G(\underline{\mathbf{D}})
\end{align*}
as desired.

Finally consider the general case where $\mathbf{0}\prec\mathbf{D}_{\mathcal{S}_{k,i}} \preceq \mathbf{\Sigma}_{X}$, $k=1,\cdots,L; i=1,\cdots,2^{k-1}$. Define $\underline{\mathbf{D}}(\epsilon)=(\mathbf{D}_{\mathcal{S}_{k,i}}-\epsilon\mathbf{I}_m)_{k=1,\cdots,L;i=1,\cdots,2^{k-1}}$. The preceding argument implies
\begin{align}
\liminf\limits_{\epsilon\downarrow 0}R_G(\underline{\mathbf{D}}(\epsilon))=\liminf\limits_{\epsilon\downarrow 0}R^*_G(\underline{\mathbf{D}}(\epsilon)). \label{eq:imply1}
\end{align}
Since $R_G(\underline{\mathbf{D}})\leq R_G(\underline{\mathbf{D}}(\epsilon))$ for any $\epsilon>0$, we have
\begin{align}
R_G(\underline{\mathbf{D}})\leq\liminf\limits_{\epsilon\downarrow 0}R_G(\underline{\mathbf{D}}(\epsilon)).\label{eq:imply2}
\end{align}
Moreover, it follows by the continuity of $R_G(\underline{\mathbf{D}}(\epsilon))$ at $\epsilon=0$ that
\begin{align}
\liminf\limits_{\epsilon\downarrow 0}R^*_G(\underline{\mathbf{D}}(\epsilon))=R^*_G(\underline{\mathbf{D}}).\label{eq:imply3}
\end{align}
Combining \eqref{eq:imply1}--\eqref{eq:imply3} completes the proof.

\emph{Remark:} Now we are in a position to make a comparison with the converse argument in Appendix \ref{LB}. Roughly speaking, $U_j$ corresponds to $\varphi_{j}^{n}(X^{n})$, $j=1,\cdots,M$, and $\tilde{X}_{{k,i}}$ can be viewed as the single-letter version of $\mathbb{E}[X^n|Z^{n}_{k,i}]$ (assuming $X^n$ and $Z^{n}_{k,i}$ are jointly Gaussian), $k=1,\cdots,L-1;i=1,\cdots,2^{k-1}$. In this converse argument, we obtain a lower bound by dropping the terms $I((\varphi_{j}^{n}(X^{n}))_{j \in \mathcal{S}_{k+1, 2i-1}};(\varphi_{j}^{(n)}(X^{n}))_{j \in \mathcal{S}_{k+1, 2i}  }|Z^{n}_{k,i})$, $k=1,\cdots,L-1;i=1,\cdots,2^{k-1}$ (see (\ref{eq:compareconverse})). For this lower bound to be attainable, one must ensure that the corresponding terms $I((U_j)_{j\in\mathcal{S}_{k+1,2i-1}} ;(U_j)_{j\in\mathcal{S}_{k+1,2i}}|\tilde{X}_{{k,i}})$, $k=1,\cdots,L-1;i=1,\cdots,2^{k-1}$, in the achievability part can be made equal to zero. Although this is not always possible for a generic version of the extended El Gamal-Cover scheme, as we have shown (see (\ref{eq:keyindep})), the specific construction based on $\underline{\mathbf{\Theta}}^*$ and $\underline{\tilde{\mathbf{\Sigma}}}$ indeed possesses the desired conditional independence structures, which can be effectively exploited by Gauss-Markov tree-structured auxiliary random variables to establish a matching converse.

\section{Conclusion}\label{sec:conclusion}
We have generalized Ozarow's converse argument to establish a single-letter lower bound on the sum rate of multiple description coding with tree-structured distortion constraints. Moreover, this lower bound is shown to be tight for the quadratic vector Gaussian case. It is worth mentioning that the applicability of Ozarow's method is by no means confined to the setting of tree-structured distortion constraints considered in the present work. Indeed, this method has been successfully used to obtain conclusive results for the multiple description problem with symmetrical distortion constraints \cite{WV09,SSC14}. Here it is curious to observe that the projection of the entropy region to the coordinates for which the corresponding subsets form a tree structure and the projected entropy region induced by the symmetric group \cite{CSLT16} are both completely characterized by the Shannon type inequalities. This suggests that the difficulty in solving the general $M$-description problem is potentially related to the lack of explicit characterization of the entropy region, and it might be more fruitful to focus on those special formulations of the multiple description problem for which the corresponding projected entropy regions admit simple characterizations.


\appendices

\section{Proof of Theorem \ref{lemma:LB}} \label{LB}
Consider arbitrary encoding functions
\begin{align*}
\varphi_{j}^{(n)}: \mathcal{X}^{n} \rightarrow \mathcal{C}_{j}^{(n)}, \quad j=1,\cdots,2^{L-1},
\end{align*}
and decoding functions
\begin{align*}
\psi_{\mathcal{S}_{k,i}}^{(n)}: \prod_{j\in\mathcal{S}_{k,i}}\mathcal{C}_{j}^{(n)} \rightarrow \hat{\mathcal{X}}^{n},\quad k=1,\cdots,L; i=1,\cdots,2^{k-1},
\end{align*}
satisfying
\begin{align*}
\frac{1}{n} \sum_{t=1}^{n} \mathbb{E} [ d(X(t), \hat{X}_{\mathcal{S}_{k,i}}(t)) ] \leq d_{\mathcal{S}_{k,i}}, \quad k=1,\cdots,L; i=1,\cdots,2^{k-1}.
\end{align*}
Note that
\begin{align}
\sum_{j=1}^{2^{L-1}} \log |\mathcal{C}_{j}^{(n)}|&\geq \sum_{j=1}^{2^{L-1}} H(\varphi_{j}^{(n)}(X^{n})) \nonumber  \\
&= \sum_{i=1}^{2^{L-1}}H((\varphi_{j}^{(n)}(X^{n}))_{j \in \mathcal{S}_{L, i}} ). \label{eq:sumc}
\end{align}
Successively applying the fact that
\begin{align*}
&\sum\limits_{i=1}^{2^{k}}H((\varphi_{j}^{(n)}(X^{n}))_{j \in \mathcal{S}_{k+1, i}} )\nonumber\\
&=\sum\limits_{i=1}^{2^{k-1}}H((\varphi_{j}^{(n)}(X^{n}))_{j \in \mathcal{S}_{k, i}} )+\sum\limits_{i=1}^{2^{k-1}}I((\varphi_{j}^{(n)}(X^{n}))_{j \in \mathcal{S}_{k+1, 2i-1}};(\varphi_{j}^{(n)}(X^{n}))_{j \in \mathcal{S}_{k+1, 2i}} )
\end{align*}
from $k=L-1$ to $k=1$ shows
\begin{align}
&\sum_{i=1}^{2^{L-1}}H((\varphi_{j}^{(n)}(X^{n}))_{j \in \mathcal{S}_{L, i}} )\nonumber\\
&=H((\varphi_{j}^{(n)}(X^{n}))_{j \in \mathcal{S}_{1, 1}}) + \sum_{k=1}^{{L-1}} \sum_{i=1}^{2^{k-1}}I((\varphi_{j}^{(n)}(X^{n}))_{j \in \mathcal{S}_{k+1, 2i-1}}; (\varphi_{j}^{(n)}(X^{n}))_{ j \in \mathcal{S}_{k+1, 2i}}).\label{eq:succ}
\end{align}
Substituting (\ref{eq:succ}) into (\ref{eq:sumc}) and invoking the fact that $H((\varphi_{j}^{(n)}(X^{n}))_{j \in \mathcal{S}_{1, 1}})=I(X^n;(\varphi_{j}^{(n)}(X^{n}))_{j \in \mathcal{S}_{1, 1}})$ gives
\begin{align}
\sum_{j=1}^{2^{L-1}} \log |\mathcal{C}_{j}^{(n)}|\geq I(X^n;(\varphi_{j}^{(n)}(X^{n}))_{j \in \mathcal{S}_{1, 1}})+ \sum_{k=1}^{{L-1}} \sum_{i=1}^{2^{k-1}}I((\varphi_{j}^{(n)}(X^{n}))_{j \in \mathcal{S}_{k+1, 2i-1}}; (\varphi_{j}^{(n)}(X^{n}))_{ j \in \mathcal{S}_{k+1, 2i}}).\label{eqn:bad}
\end{align}
Now introduce an auxiliary remote vector source $\{(Z_{k,i}(t))_{k =1,\cdots, L-1; i =1, \cdots, 2^{k-1}}\}_{t=1}^{\infty}$, which forms a joint i.i.d. process with $\{X(t)\}_{t=1}^\infty$. We assume that the conditional distribution of $(Z_{k,i}(t))_{k =1,\cdots, L-1; i =1, \cdots, 2^{k-1}}$ given $X(t)$ satisfies the binary Markov tree condition (\ref{eq:Markovtree}) for every $t$.
It can be verified that
\begin{align}
& I((\varphi_{j}^{(n)}(X^{n}))_{j \in \mathcal{S}_{k+1, 2i-1}};  (\varphi_{j}^{n}(X^{n}))_{j \in \mathcal{S}_{k+1, 2i}}) \nonumber \\
&= I(Z^{n}_{k,i}; (\varphi_{j}^{n}(X^{n}))_{j \in \mathcal{S}_{k+1, 2i-1}}) \nonumber \\
&\quad + I(Z^{n}_{k,i}; (\varphi_{j}^{(n)}(X^{n}))_{j \in \mathcal{S}_{k+1, 2i}}) \nonumber \\
&\quad - I(Z^{n}_{k,i}; (\varphi_{j}^{(n)}(X^{n}))_{j \in \mathcal{S}_{k, i}}) \nonumber \\
&\quad +  I((\varphi_{j}^{n}(X^{n}))_{j \in \mathcal{S}_{k+1, 2i-1}};(\varphi_{j}^{(n)}(X^{n}))_{j \in \mathcal{S}_{k+1, 2i}  }|Z^{n}_{k,i})\nonumber\\
&\geq  I(Z^{n}_{k,i}; (\varphi_{j}^{(n)}(X^{n}))_{j \in \mathcal{S}_{k+1, 2i-1}}) \nonumber \\
&\quad + I(Z^{n}_{k,i}; (\varphi_{j}^{(n)}(X^{n}))_{j \in \mathcal{S}_{k+1, 2i}}) \nonumber \\
&\quad - I(Z^{n}_{k,i}; (\varphi_{j}^{n}(X^{n}))_{j \in \mathcal{S}_{k, i}}),\quad k =1,\cdots, L-1; i =1, \cdots, 2^{k-1}.\label{eq:compareconverse}
\end{align}
We have
\begin{align}
&\sum_{k=1}^{{L-1}} \sum_{i=1}^{2^{k-1}}I((\varphi_{j}^{(n)}(X^{n}))_{j \in \mathcal{S}_{k+1, 2i-1}};  (\varphi_{j}^{n}(X^{n}))_{j \in \mathcal{S}_{k+1, 2i}})\nonumber\\
&\geq\sum_{k=1}^{{L-1}} \sum_{i=1}^{2^{k-1}}(I(Z^{n}_{k,i}; (\varphi_{j}^{(n)}(X^{n}))_{j \in \mathcal{S}_{k+1, 2i-1}})+I(Z^{n}_{k,i}; (\varphi_{j}^{(n)}(X^{n}))_{j \in \mathcal{S}_{k+1, 2i}}))\nonumber\\
&\quad-\sum_{k=1}^{{L-1}} \sum_{i=1}^{2^{k-1}}I(Z^{n}_{k,i}; (\varphi_{j}^{n}(X^{n}))_{j \in \mathcal{S}_{k, i}})\nonumber\\
&=-I(Z_{1,1}^{n}; (\varphi_{j}^{(n)}(X^{n}))_{j \in \mathcal{S}_{1, 1}})+\sum_{k=1}^{{L-1}} \sum_{i=1}^{2^{k-1}}(I(Z^{n}_{k,i}; (\varphi_{j}^{(n)}(X^{n}))_{j \in \mathcal{S}_{k+1, 2i-1}})+I(Z^{n}_{k,i}; (\varphi_{j}^{(n)}(X^{n}))_{j \in \mathcal{S}_{k+1, 2i}}))\nonumber\\
&\quad-\sum_{k=1}^{{L-2}} \sum_{i=1}^{2^{k-1}}(I(Z^{n}_{k+1,2i-1}; (\varphi_{j}^{n}(X^{n}))_{j \in \mathcal{S}_{k+1, 2i-1}})+I(Z^{n}_{k+1,2i}; (\varphi_{j}^{n}(X^{n}))_{j \in \mathcal{S}_{k+1, 2i}}))\nonumber\\
&=-I(Z_{1,1}^{n}; (\varphi_{j}^{(n)}(X^{n}))_{j \in \mathcal{S}_{1, 1}})\nonumber\\
&\quad+\sum_{k=1}^{{L-2}} \sum_{i=1}^{2^{k-1}}(I(Z^{n}_{k,i}; (\varphi_{j}^{(n)}(X^{n}))_{j \in \mathcal{S}_{k+1, 2i-1}})-I(Z^{n}_{k+1,2i-1}; (\varphi_{j}^{n}(X^{n}))_{j \in \mathcal{S}_{k+1, 2i-1}})\nonumber\\
&\hspace{0.8in}+I(Z^{n}_{k,i}; (\varphi_{j}^{(n)}(X^{n}))_{j \in \mathcal{S}_{k+1, 2i}})-I(Z^{n}_{k+1,2i}; (\varphi_{j}^{n}(X^{n}))_{j \in \mathcal{S}_{k+1, 2i}}))\nonumber\\
&\quad+\sum_{i=1}^{2^{L-2}}(I(Z^{n}_{L-1,i}; (\varphi_{j}^{(n)}(X^{n}))_{j \in \mathcal{S}_{L, 2i-1}})+I(Z^{n}_{L-1,i}; (\varphi_{j}^{(n)}(X^{n}))_{j \in \mathcal{S}_{L, 2i}})).
\label{eqn:bd}
\end{align}
Substituting \eqref{eqn:bd} into \eqref{eqn:bad} gives
\begin{align}
\sum_{j=1}^{2^{L-1}} \log |\mathcal{C}_{j}^{(n)}| &\geq I(X^{n}; (\varphi_{j}^{(n)}(X^{n}))_{j \in \mathcal{S}_{1, 1}})-I(Z_{1,1}^{n}; (\varphi_{j}^{(n)}(X^{n}))_{j \in \mathcal{S}_{1, 1}})\nonumber\\
&\quad+\sum_{k=1}^{{L-2}} \sum_{i=1}^{2^{k-1}}(I(Z^{n}_{k,i}; (\varphi_{j}^{(n)}(X^{n}))_{j \in \mathcal{S}_{k+1, 2i-1}})-I(Z^{n}_{k+1,2i-1}; (\varphi_{j}^{n}(X^{n}))_{j \in \mathcal{S}_{k+1, 2i-1}})\nonumber\\
&\hspace{0.8in}+I(Z^{n}_{k,i}; (\varphi_{j}^{(n)}(X^{n}))_{j \in \mathcal{S}_{k+1, 2i}})-I(Z^{n}_{k+1,2i}; (\varphi_{j}^{n}(X^{n}))_{j \in \mathcal{S}_{k+1, 2i}}))\nonumber\\
&\quad+\sum_{i=1}^{2^{L-2}}(I(Z^{n}_{L-1,i}; (\varphi_{j}^{(n)}(X^{n}))_{j \in \mathcal{S}_{L, 2i-1}})+I(Z^{n}_{L-1,i}; (\varphi_{j}^{(n)}(X^{n}))_{j \in \mathcal{S}_{L, 2i}})).\label{eq:keyexp}
\end{align}
Let $T$ be a random variable uniformly distributed over $\{1,\cdots,n\}$ and independent of everything else. We have
\begin{align}
& I(X^{n}; (\varphi_{j}^{(n)}(X^{n}))_{j \in \mathcal{S}_{1, 1}})   -I(Z_{1,1}^{n}; (\varphi_{j}^{(n)}(X^{n}))_{j \in \mathcal{S}_{1, 1}} ) \nonumber \\
&=  I(Z_{1,1}^{n}, X^{n}; (\varphi_{j}^{(n)}(X^{n}))_{j \in \mathcal{S}_{1, 1}})  -I(Z_{1,1}^{n}; (\varphi_{j}^{(n)}(X^{n}))_{j \in \mathcal{S}_{1, 1}} )\nonumber  \\
&=  I( X^{n};(\varphi_{j}^{(n)}(X^{n}))_{j \in \mathcal{S}_{1, 1} }| Z_{1,1}^{n})\nonumber \\
&\geq  I( X^{n};\hat{X}_{\mathcal{S}_{1, 1}}^{n} | Z_{1,1}^{n})  \nonumber\\
&=   \sum_{t=1}^{n} I(X(t); \hat{X}_{\mathcal{S}_{1, 1}}^{n} | X(1:t-1), Z_{1,1}^{n})\nonumber  \\
&=   \sum_{t=1}^{n} I(X(t); \hat{X}_{\mathcal{S}_{1, 1}}^{n},X(1:t-1),Z_{1,1}(1:t-1),Z_{1,1}(t+1:n) | Z_{1,1}(t)) \nonumber \\
&\geq   \sum_{t=1}^{n} I (X(t); \hat{X}_{\mathcal{S}_{1, 1}}(t) | Z_{1,1}(t)) \nonumber\\
&=   n I (X(T); \hat{X}_{\mathcal{S}_{1, 1}}(T) | Z_{1,1}(T),T) \nonumber\\
&=   n I (X(T); \hat{X}_{\mathcal{S}_{1, 1}}(T),T | Z_{1,1}(T)) \nonumber\\
&\geq  n I (X(T); \hat{X}_{\mathcal{S}_{1, 1}}(T) | Z_{1,1}(T)).\label{eqn:ter1}
\end{align}
Similarly, it can be shown that
\begin{align}
& I(Z^{n}_{k,i};(\varphi_{j}^{(n)}(X^{n}))_{j \in \mathcal{S}_{k+1, 2i-1}} ) -I(Z^{n}_{k+1, 2i-1}; (\varphi_{j}^{(n)}(X^{n}))_{j \in \mathcal{S}_{k+1, 2i-1}} ) \nonumber \\
 &\geq   n I(Z_{k,i}(T); \hat{X}_{\mathcal{S}_{k+1,2i-1}}(T) | Z_{k+1,2i-1}(T)), \quad k=1,\cdots,L-2; i=1,\cdots,2^{k-1}, \label{eqn:ter2} \\
 &  I(Z^{n}_{k,i};(\varphi_{j}^{(n)}(X^{n}))_{j \in \mathcal{S}_{k+1, 2i}} ) -I(Z^{n}_{k+1,2i}; (\varphi_{j}^{(n)}(X^{n}))_{j \in \mathcal{S}_{k+1, 2i}} ) \nonumber \\
  &\geq   n I(Z_{k,i}(T);  \hat{X}_{\mathcal{S}_{k+1,2i}}(T) | Z_{k+1, 2i}(T)),\quad k=1,\cdots,L-2; i=1,\cdots,2^{k-1}, \label{eqn:ter3} \\
 & I (Z_{L-1,i}^{n}; (\varphi_{j}^{(n)}(X^{n}))_{j \in \mathcal{S}_{L, 2i-1}})\geq  n I (Z_{L-1,i}(T); \hat{X}_{\mathcal{S}_{L ,2i-1}}(T)),\quad i=1,\cdots,2^{L-2}, \label{eqn:ter4} \\
  & I (Z_{L-1,i}^{n}; (\varphi_{j}^{(n)}(X^{n}))_{j \in \mathcal{S}_{L, 2i}})\geq  n I (Z_{L-1,i}(T); \hat{X}_{\mathcal{S}_{L ,2i}}(T)),\quad i=1,\cdots,2^{L-2}. \label{eqn:ter5}
\end{align}
Substituting \eqref{eqn:ter1}--\eqref{eqn:ter5} into (\ref{eq:keyexp}) gives
\begin{align}
\sum_{j=1}^{2^{L-1}} \log |\mathcal{C}_{j}^{(n)}| &\geq n I(X(T); \hat{X}_{\mathcal{S}_{1,1}}(T)|Z_{1,1}(T))\nonumber \\
&\quad+ \sum_{k=1}^{L-2} \sum_{i=1}^{2^{k-1}} (n I(Z_{k,i}(T); \hat{X}_{\mathcal{S}_{k+1,2i-1}}(T) | Z_{k+1,2i-1}(T)) \nonumber \\
 &\hspace{0.8in} + n I(Z_{k,i}(T);  \hat{X}_{\mathcal{S}_{k+1,2i}}(T) | Z_{k+1,2i}(T)))  \nonumber \\
 &\quad+ \sum_{i=1}^{2^{L-2}} ( n I (Z_{L-1,i}(T); \hat{X}_{\mathcal{S}_{L ,2i-1}}(T))+ n  I (Z_{L-1,i}(T); \hat{X}_{\mathcal{S}_{L ,2i}}(T)).
\end{align}
Note that $(Z_{k,i}(T))_{k =1,\cdots, L-1; i =1, \cdots, 2^{k-1}}\leftrightarrow X(T)\leftrightarrow (\hat{X}_{\mathcal{S}_{k,i}}(T))_{k =1,\cdots, L; i =1, \cdots, 2^{k-1}}$ form a Markov chain; moreover, the conditional distribution of $(Z_{k,i}(T))_{k =1,\cdots, L-1; i =1, \cdots, 2^{k-1}}$ given $X(T)$ is in $\mathcal{P}$ while the conditional distribution of $(\hat{X}_{\mathcal{S}_{k,i}}(T))_{k =1,\cdots, L; i =1, \cdots, 2^{k-1}}$ given $X(T)$ is in $\mathcal{P}(\underline{d})$. This completes the proof of Theorem \ref{lemma:LB}.

\section{Proof of Lemma \ref{lemma:KKT}}\label{app:KKT}

The Lagrangian of the maximization problem in (\ref{eq:mainequiv}) is given by
\begin{align}
 \mathcal{L}&= \sum_{k=1}^{L-1} \sum_{i=1}^{2^{k-1}}(\log | \mathbf{\Theta}_{k,i}+ \mathbf{\Sigma}_{\mathcal{S}_{k,i}}| -\log | \mathbf{\Theta}_{k,i}+ \mathbf{\Sigma}_{\mathcal{S}_{k+1,2i-1}} |
-\log | \mathbf{\Theta}_{k,i}+ \mathbf{\Sigma}_{\mathcal{S}_{k+1,2i}}| )  \nonumber \\
&\quad+ \tr{(  \mathbf{\Theta}_{1,1} \mathbf{M}_{1,1})} + \sum_{k=1}^{L-2} \sum_{i=1}^{2^{k-1}} ( \tr{((  \mathbf{\Theta}_{k+1,2i-1}-    \mathbf{\Theta}_{k,i}    )\mathbf{M}_{k+1,2i-1} )} + \tr{((  \mathbf{\Theta}_{k+1,2i}-    \mathbf{\Theta}_{k,i}    )\mathbf{M}_{k+1,2i} )} ) \nonumber \\
&\quad+ \sum_{i=1}^{2^{L-2}}\tr ((  \mathbf{\Sigma}_{X}  -   \mathbf{\Theta}_{L-1, i}) \mathbf{M}_{L,2i-1}),\nonumber
\end{align}
where $\mathbf{M}_{k,i}$, $k=1,\cdots,L; i=1,\cdots,2^{k-1}$, are the Lagrange multipliers\footnote{Note that $\mathbf{M}_{L,2i}$, $i=1,\cdots,2^{L-2}$, do not appear in the Lagrangian. They are introduced to simplify notation, and are set to zero.} satisfying
\begin{align*}
&\mathbf{M}_{k,i}  \succeq \mathbf{0}, \quad k=1, \cdots, L-1; i=1, \cdots, 2^{k-1}, \\
& \mathbf{M}_{L,2i-1} \succeq \mathbf{0}, \quad i=1,\cdots,2^{L-2}, \\
& \mathbf{M}_{L,2i} = \mathbf{0}, \quad i=1,\cdots,2^{L-2}.
\end{align*}
It can be shown by leveraging \cite[Proposition 3.3.11, p. 327]{NP} that, for any optimal solution $\underline{\mathbf{\Theta}}^*$, there exist $\underline{\mathbf{M}}^*\triangleq(\mathbf{M}^*_{k,i})_{k=1,\cdots,L; i=1,\cdots,2^{k-1}}$, such that
\begin{align}
&\left.\nabla_{\mathbf{\Theta}_{k,i}}\mathcal{L}\right|_{\underline{\mathbf{\Theta}}=\underline{\mathbf{\Theta}}^*,\underline{\mathbf{M}}=\underline{\mathbf{M}}^*}=\mathbf{0},\quad k=1,\cdots,L-1;i=1,\cdots,2^{k-1},\label{eq:newKKT}\\
& \mathbf{M}^*_{1,1}\mathbf{\Theta}_{1,1}^{*} = \mathbf{\Theta}_{1,1}^{*}\mathbf{M}^*_{1,1}  = \mathbf{0}, \nonumber\\
& \mathbf{M}^*_{k+1,2i-1} (\mathbf{\Theta}_{k+1,2i-1}^{*} - \mathbf{\Theta}_{k,i}^{*}) =  (\mathbf{\Theta}_{k+1,2i-1}^{*} - \mathbf{\Theta}_{k,i}^{*}) \mathbf{M}^*_{k+1,2i-1}  = \mathbf{0}, \nonumber\\
&\hspace{2in} k=1,\cdots,L-2; i=1,\cdots,2^{k-1}, \nonumber\\
& \mathbf{M}^*_{k+1,2i} (\mathbf{\Theta}_{k+1,2i}^{*} - \mathbf{\Theta}_{k,i}^{*}) =  (\mathbf{\Theta}_{k+1,2i}^{*} - \mathbf{\Theta}_{k,i}^{*}) \mathbf{M}^*_{k+1,2i}  = \mathbf{0}, \nonumber\\
&\hspace{2in} k=1,\cdots,L-2; i=1,\cdots,2^{k-1}, \nonumber \\
& \mathbf{M}^*_{L,2i-1} (\mathbf{\Sigma}_{X} - \mathbf{\Theta}_{L-1,i}^{*}) =  (\mathbf{\Sigma}_{X} - \mathbf{\Theta}_{L-1,i}^{*}) \mathbf{M}^*_{L,2i-1}  = \mathbf{0}, \quad i=1,\cdots,2^{L-2}, \nonumber\\
&\mathbf{M}^*_{k,i}  \succeq \mathbf{0}, \quad k=1, \cdots, L-1; i=1, \cdots, 2^{k-1},\nonumber\\
& \mathbf{M}^*_{L,2i-1} \succeq \mathbf{0}, \quad i=1,\cdots,2^{L-2}, \nonumber\\
& \mathbf{M}^*_{L,2i} = \mathbf{0}, \quad i=1,\cdots,2^{L-2}.\nonumber
\end{align}
Note that
\begin{align*}
\mathcal{L}&= \sum_{k=1}^{L-1} \sum_{i=1}^{2^{k-1}}(\log | \mathbf{\Theta}_{k,i}+ \mathbf{\Sigma}_{\mathcal{S}_{k,i}}| -\log | \mathbf{\Theta}_{k,i}+ \mathbf{\Sigma}_{\mathcal{S}_{k+1,2i-1}} |
-\log | \mathbf{\Theta}_{k,i}+ \mathbf{\Sigma}_{\mathcal{S}_{k+1,2i}}|)  \nonumber \\
&\quad+ \tr{(  \mathbf{\Theta}_{1,1} \mathbf{M}_{1,1})} + \sum_{k=1}^{L-2} \sum_{i=1}^{2^{k-1}} ( \tr{( \mathbf{\Theta}_{k+1,2i-1}\mathbf{M}_{k+1,2i-1} )} + \tr{(  \mathbf{\Theta}_{k+1,2i}   \mathbf{M}_{k+1,2i} )} ) \nonumber \\
&\quad-\sum_{k=1}^{L-2} \sum_{i=1}^{2^{k-1}} \tr(\mathbf{\Theta}_{k,i}(\mathbf{M}_{k+1,2i-1}+\mathbf{M}_{k+1,2i}))-\sum\limits_{i=1}^{2^{L-2}}\tr(\mathbf{\Theta}_{L-1, i}\mathbf{M}_{L,2i-1})+ \sum_{i=1}^{2^{L-2}}\tr (\mathbf{\Sigma}_{X}\mathbf{M}_{L,2i-1})\\
&= \sum_{k=1}^{L-1} \sum_{i=1}^{2^{k-1}}(\log | \mathbf{\Theta}_{k,i}+ \mathbf{\Sigma}_{\mathcal{S}_{k,i}}| -\log | \mathbf{\Theta}_{k,i}+ \mathbf{\Sigma}_{\mathcal{S}_{k+1,2i-1}} |
-\log | \mathbf{\Theta}_{k,i}+ \mathbf{\Sigma}_{\mathcal{S}_{k+1,2i}}| )  \nonumber \\
&\quad+ \sum_{k=1}^{L-1} \sum_{i=1}^{2^{k-1}} \tr(\mathbf{\Theta}_{k,i}\mathbf{M}_{k,i})-\sum_{k=1}^{L-1} \sum_{i=1}^{2^{k-1}} \tr(\mathbf{\Theta}_{k,i}(\mathbf{M}_{k+1,2i-1}+\mathbf{M}_{k+1,2i})) \nonumber \\
&\quad+ \sum_{i=1}^{2^{L-2}}\tr (\mathbf{\Sigma}_{X}\mathbf{M}_{L,2i-1})\nonumber\\
&=\sum_{k=1}^{L-1} \sum_{i=1}^{2^{k-1}}(\log | \mathbf{\Theta}_{k,i}+ \mathbf{\Sigma}_{\mathcal{S}_{k,i}}| -\log | \mathbf{\Theta}_{k,i}+ \mathbf{\Sigma}_{\mathcal{S}_{k+1,2i-1}} |
-\log | \mathbf{\Theta}_{k,i}+ \mathbf{\Sigma}_{\mathcal{S}_{k+1,2i}}|)  \nonumber \\
&\quad+ \sum_{k=1}^{L-1} \sum_{i=1}^{2^{k-1}} \tr(\mathbf{\Theta}_{k,i}(\mathbf{M}_{k,i}-\mathbf{M}_{k+1,2i-1}-\mathbf{M}_{k+1,2i})) \nonumber \\
&\quad+ \sum_{i=1}^{2^{L-2}}\tr (\mathbf{\Sigma}_{X}\mathbf{M}_{L,2i-1}).
\end{align*}
Therefore, we have
\begin{align}
\left.\nabla_{\mathbf{\Theta}_{k,i}}\mathcal{L}\right|_{\underline{\mathbf{\Theta}}=\underline{\mathbf{\Theta}}^*,\underline{\mathbf{M}}=\underline{\mathbf{M}}^*}&=(\mathbf{\Theta}^*_{k,i}+ \mathbf{\Sigma}_{\mathcal{S}_{k,i}})^{-1}-(\mathbf{\Theta}^*_{k,i}+ \mathbf{\Sigma}_{\mathcal{S}_{k+1,2i-1}})^{-1}-(\mathbf{\Theta}^*_{k,i}+ \mathbf{\Sigma}_{\mathcal{S}_{k+1,2i}})^{-1}\nonumber\\
&\quad+\mathbf{M}^*_{k,i}-\mathbf{M}^*_{k+1,2i-1}-\mathbf{M}^*_{k+1,2i},\quad k=1,\cdots,L-1;i=1,\cdots,2^{k-1}.\label{eq:lemmasub}
\end{align}
Substituting (\ref{eq:lemmasub}) into (\ref{eq:newKKT}) completes the proof of Lemma \ref{lemma:KKT}.

\section{Proof of Lemma \ref{lemma:enhance}} \label{enhance}

Note that
\begin{align}
 & \mathbf{\Theta}_{k,i}^{*} +\tilde{\mathbf{\Sigma}}_{\mathcal{S}_{k+1,2i-1}} \nonumber \\
 &= \mathbf{\Theta}_{k+1,2i-1}^{*} +\tilde{\mathbf{\Sigma}}_{\mathcal{S}_{k+1,2i-1}} - (\mathbf{\Theta}_{k+1,2i-1}^{*}- \mathbf{\Theta}_{k,i}^{*} )\nonumber \\
 & = (( \mathbf{\Theta}_{k+1,2i-1}^{*} +{\mathbf{\Sigma}}_{\mathcal{S}_{k+1,2i-1}} )^{-1} + \mathbf{M}^*_{k+1,2i-1})^{-1}- (\mathbf{\Theta}_{k+1,2i-1}^{*}- \mathbf{\Theta}_{k,i}^{*} ) \label{eq:duedef}\\
 &= ( \mathbf{I}_m + ( \mathbf{\Theta}_{k+1,2i-1}^{*} +{\mathbf{\Sigma}}_{\mathcal{S}_{k+1,2i-1}} ) \mathbf{M}^*_{k+1,2i-1})^{-1} ( \mathbf{\Theta}_{k+1,2i-1}^{*} +{\mathbf{\Sigma}}_{\mathcal{S}_{k+1,2i-1}} )- (\mathbf{\Theta}_{k+1,2i-1}^{*}- \mathbf{\Theta}_{k,i}^{*} ) \nonumber\\
 & =  ( \mathbf{I}_m + ( \mathbf{\Theta}_{k,i}^{*} +{\mathbf{\Sigma}}_{\mathcal{S}_{k+1,2i-1}} ) \mathbf{M}^*_{k+1,2i-1})^{-1} ( \mathbf{\Theta}_{k+1,2i-1}^{*} +{\mathbf{\Sigma}}_{\mathcal{S}_{k+1,2i-1}} )- (\mathbf{\Theta}_{k+1,2i-1}^{*}- \mathbf{\Theta}_{k,i}^{*} )\label{eq:dueKKT1} \\
  &=  ( ( \mathbf{\Theta}_{k,i}^{*} +{\mathbf{\Sigma}}_{\mathcal{S}_{k+1,2i-1}} )^{-1} + \mathbf{M}^*_{k+1,2i-1} )^{-1}( \mathbf{\Theta}_{k,i}^{*} +{\mathbf{\Sigma}}_{\mathcal{S}_{k+1,2i-1}} )^{-1}( \mathbf{\Theta}_{k+1,2i-1}^{*} +{\mathbf{\Sigma}}_{\mathcal{S}_{k+1,2i-1}} )\nonumber \\
  & \quad - (\mathbf{\Theta}_{k+1,2i-1}^{*}- \mathbf{\Theta}_{k,i}^{*} ) \nonumber\\
  &=  (( \mathbf{\Theta}_{k,i}^{*} +{\mathbf{\Sigma}}_{\mathcal{S}_{k+1,2i-1}} )^{-1} + \mathbf{M}^*_{k+1,2i-1} )^{-1}( \mathbf{\Theta}_{k,i}^{*} +{\mathbf{\Sigma}}_{\mathcal{S}_{k+1,2i-1}} )^{-1}( \mathbf{\Theta}_{k+1,2i-1}^{*} -{\mathbf{\Theta}}_{k,i}^{*} )\nonumber \\
  & \quad +(( \mathbf{\Theta}_{k,i}^{*} +{\mathbf{\Sigma}}_{\mathcal{S}_{k+1,2i-1}} )^{-1} + \mathbf{M}^*_{k+1,2i-1} )^{-1} - (\mathbf{\Theta}_{k+1,2i-1}^{*}- \mathbf{\Theta}_{k,i}^{*} ) \nonumber\\
  &=  (( \mathbf{\Theta}_{k,i}^{*} +{\mathbf{\Sigma}}_{\mathcal{S}_{k+1,2i-1}} )^{-1} + \mathbf{M}^*_{k+1,2i-1} )^{-1}(( \mathbf{\Theta}_{k,i}^{*} +{\mathbf{\Sigma}}_{\mathcal{S}_{k+1,2i-1}} )^{-1}+ \mathbf{M}^*_{k+1,2i-1} )( \mathbf{\Theta}_{k+1,2i-1}^{*} -{\mathbf{\Theta}}_{k,i}^{*} )\nonumber \\
  & \quad +( ( \mathbf{\Theta}_{k,i}^{*} +{\mathbf{\Sigma}}_{\mathcal{S}_{k+1,2i-1}} )^{-1} + \mathbf{M}^*_{k+1,2i-1} )^{-1} - (\mathbf{\Theta}_{k+1,2i-1}^{*}- \mathbf{\Theta}_{k,i}^{*} ) \label{eq:dueKKT2}\\
  &=  ( \mathbf{\Theta}_{k+1,2i-1}^{*} -{\mathbf{\Theta}}_{k,i}^{*} ) +( ( \mathbf{\Theta}_{k,i}^{*} +{\mathbf{\Sigma}}_{\mathcal{S}_{k+1,2i-1}} )^{-1} + \mathbf{M}^*_{k+1,2i-1} )^{-1} - (\mathbf{\Theta}_{k+1,2i-1}^{*}- \mathbf{\Theta}_{k,i}^{*} )\nonumber\\
  &= ( ( \mathbf{\Theta}_{k,i}^{*} +{\mathbf{\Sigma}}_{\mathcal{S}_{k+1,2i-1}} )^{-1} + \mathbf{M}^*_{k+1,2i-1} )^{-1},\quad k=1,\cdots, L-2; i=1, \cdots, 2^{k-1},\label{eq:appcomb1}
\end{align}
where (\ref{eq:duedef}) is due to \eqref{eq:en1} while (\ref{eq:dueKKT1}) and (\ref{eq:dueKKT2}) are due to \eqref{eq:KKT3}. Moreover,
\begin{align}
 & \mathbf{\Theta}_{L-1,i}^{*} +\tilde{\mathbf{\Sigma}}_{\mathcal{S}_{L,2i-1}} \nonumber \\
 &= \mathbf{\Sigma}_X +\tilde{\mathbf{\Sigma}}_{\mathcal{S}_{L,2i-1}} - (\mathbf{\Sigma}_X- \mathbf{\Theta}_{L-1,i}^{*} )\nonumber \\
 & = (( \mathbf{\Sigma}_X +{\mathbf{\Sigma}}_{\mathcal{S}_{L,2i-1}} )^{-1} + \mathbf{M}^*_{L,2i-1})^{-1}- (\mathbf{\Sigma}_X- \mathbf{\Theta}_{L-1,i}^{*} ) \label{eq:2duedef}\\
 &= ( \mathbf{I}_m + ( \mathbf{\Sigma}_X +{\mathbf{\Sigma}}_{\mathcal{S}_{L,2i-1}} ) \mathbf{M}^*_{L,2i-1})^{-1} ( \mathbf{\Sigma}_X +{\mathbf{\Sigma}}_{\mathcal{S}_{L,2i-1}} )- (\mathbf{\Sigma}_X- \mathbf{\Theta}_{L-1,i}^{*} ) \nonumber\\
 & =  ( \mathbf{I}_m + ( \mathbf{\Theta}_{L-1,i}^{*} +{\mathbf{\Sigma}}_{\mathcal{S}_{L,2i-1}} ) \mathbf{M}^*_{L,2i-1})^{-1} ( \mathbf{\Sigma}_X +{\mathbf{\Sigma}}_{\mathcal{S}_{L,2i-1}} )- (\mathbf{\Sigma}_X- \mathbf{\Theta}_{L-1,i}^{*} )\label{eq:2dueKKT1} \\
  &=  ( ( \mathbf{\Theta}_{L-1,i}^{*} +{\mathbf{\Sigma}}_{\mathcal{S}_{L,2i-1}} )^{-1} + \mathbf{M}^*_{L,2i-1} )^{-1}( \mathbf{\Theta}_{L-1,i}^{*} +{\mathbf{\Sigma}}_{\mathcal{S}_{L,2i-1}} )^{-1}( \mathbf{\Sigma}_X +{\mathbf{\Sigma}}_{\mathcal{S}_{L,2i-1}} )- (\mathbf{\Sigma}_X- \mathbf{\Theta}_{L-1,i}^{*} ) \nonumber\\
  &=  (( \mathbf{\Theta}_{L-1,i}^{*} +{\mathbf{\Sigma}}_{\mathcal{S}_{L,2i-1}} )^{-1} + \mathbf{M}^*_{L,2i-1} )^{-1}( \mathbf{\Theta}_{L-1,i}^{*} +{\mathbf{\Sigma}}_{\mathcal{S}_{L,2i-1}} )^{-1}( \mathbf{\Sigma}_X -{\mathbf{\Theta}}_{L-1,i}^{*} )\nonumber \\
  & \quad +(( \mathbf{\Theta}_{L-1,i}^{*} +{\mathbf{\Sigma}}_{\mathcal{S}_{L,2i-1}} )^{-1} + \mathbf{M}^*_{L,2i-1} )^{-1} - (\mathbf{\Sigma}_X- \mathbf{\Theta}_{L-1,i}^{*} ) \nonumber\\
  &=  (( \mathbf{\Theta}_{L-1,i}^{*} +{\mathbf{\Sigma}}_{\mathcal{S}_{L,2i-1}} )^{-1} + \mathbf{M}^*_{L,2i-1} )^{-1}(( \mathbf{\Theta}_{L-1,i}^{*} +{\mathbf{\Sigma}}_{\mathcal{S}_{L,2i-1}} )^{-1}+ \mathbf{M}^*_{L,2i-1} )( \mathbf{\Sigma}_X-{\mathbf{\Theta}}_{L-1,i}^{*} )\nonumber \\
  & \quad +( ( \mathbf{\Theta}_{L-1,i}^{*} +{\mathbf{\Sigma}}_{\mathcal{S}_{L,2i-1}} )^{-1} + \mathbf{M}^*_{L,2i-1} )^{-1} - (\mathbf{\Sigma}_X- \mathbf{\Theta}_{L-1,i}^{*} ) \label{eq:2dueKKT2}\\
  &=  ( \mathbf{\Sigma}_X -{\mathbf{\Theta}}_{L-1,i}^{*} ) +( ( \mathbf{\Theta}_{L-1,i}^{*} +{\mathbf{\Sigma}}_{\mathcal{S}_{L,2i-1}} )^{-1} + \mathbf{M}^*_{L,2i-1} )^{-1} - (\mathbf{\Sigma}_X- \mathbf{\Theta}_{L-1,i}^{*} )\nonumber\\
  &= ( ( \mathbf{\Theta}_{L-1,i}^{*} +{\mathbf{\Sigma}}_{\mathcal{S}_{L,2i-1}} )^{-1} + \mathbf{M}^*_{L,2i-1} )^{-1},\quad i=1, \cdots, 2^{L-2},\label{eq:appcomb2}
\end{align}
where (\ref{eq:2duedef}) is due to \eqref{eq:en2} while (\ref{eq:2dueKKT1}) and (\ref{eq:2dueKKT2}) are due to \eqref{eq:KKT4}. Combining (\ref{eq:appcomb1}) and (\ref{eq:appcomb2}) proves (\ref{eq:enhance1}). The proof of (\ref{eq:enhance2}) is similar to that of (\ref{eq:enhance1}) and is thus omitted.

It is easy to see that (\ref{eq:KKT_en1}) is a simple consequence of (\ref{eq:KKT1}), (\ref{eq:en1}), (\ref{eq:enhance1}), and (\ref{eq:enhance2}). In view of (\ref{eq:en1}), (\ref{eq:enhance1}), and (\ref{eq:enhance2}), we have
\begin{align*}
&( \mathbf{\Theta}_{k,i}^{*} +\tilde{\mathbf{\Sigma}}_{\mathcal{S}_{k,i}})^{-1}\succ\mathbf{0},\\
&( \mathbf{\Theta}_{k,i}^{*} +\tilde{\mathbf{\Sigma}}_{\mathcal{S}_{k+1,2i-1}})^{-1}\succ\mathbf{0},\\
&( \mathbf{\Theta}_{k,i}^{*} +\tilde{\mathbf{\Sigma}}_{\mathcal{S}_{k+1,2i}})^{-1}\succ\mathbf{0}
\end{align*}
for $k=1,\cdots,L-1; i=1,\cdots,2^{k-1}$, which, together with (\ref{eq:KKT_en1}), implies
\begin{align*}
\tilde{\mathbf{\Sigma}}_{\mathcal{S}_{k+1,2i-1}}, \tilde{\mathbf{\Sigma}}_{\mathcal{S}_{k+1,2i}}\succ  \tilde{\mathbf{\Sigma}}_{\mathcal{S}_{k,i}},\quad k=1,\cdots,L-1; i=1,\cdots,2^{k-1}.
\end{align*}
Therefore, for the purpose of proving (\ref{eq:enhance8}), it suffices to show that $\tilde{\mathbf{\Sigma}}_{\mathcal{S}_{1,1}}\succ\mathbf{0}$.
Indeed,
\begin{align}
\tilde{\mathbf{\Sigma}}_{\mathcal{S}_{1,1}} &= (( \mathbf{\Theta}_{1,1}^{*} + {\mathbf{\Sigma}}_{\mathcal{S}_{1,1}})^{-1} + \mathbf{M}^*_{1,1})^{-1} - \mathbf{\Theta}_{1,1}^{*} \label{eq:3pra}\\
&= ( \mathbf{I}_m + ( \mathbf{\Theta}_{1,1}^{*} + {\mathbf{\Sigma}}_{\mathcal{S}_{1,1}}) \mathbf{M}^*_{1,1})^{-1}( \mathbf{\Theta}_{1,1}^{*} + {\mathbf{\Sigma}}_{\mathcal{S}_{1,1}}) - \mathbf{\Theta}_{1,1}^{*} \nonumber\\
&=  ( \mathbf{I}_m +  {\mathbf{\Sigma}}_{\mathcal{S}_{1,1}} \mathbf{M}^*_{1,1} )^{-1}( \mathbf{\Theta}_{1,1}^{*} + {\mathbf{\Sigma}}_{\mathcal{S}_{1,1}}) - \mathbf{\Theta}_{1,1}^{*} \label{eq:3prb}\\
&=  ( {\mathbf{\Sigma}}_{\mathcal{S}_{1,1}}^{-1} +  \mathbf{M}^*_{1,1} )^{-1}{\mathbf{\Sigma}}_{\mathcal{S}_{1,1}}^{-1} ( \mathbf{\Theta}_{1,1}^{*} + {\mathbf{\Sigma}}_{\mathcal{S}_{1,1}}) - \mathbf{\Theta}_{1,1}^{*} \nonumber\\
&=  ( {\mathbf{\Sigma}}_{\mathcal{S}_{1,1}}^{-1} +  \mathbf{M}^*_{1,1} )^{-1} + ( {\mathbf{\Sigma}}_{\mathcal{S}_{1,1}}^{-1} +  \mathbf{M}^*_{1,1} )^{-1}{\mathbf{\Sigma}}_{\mathcal{S}_{1,1}}^{-1}  \mathbf{\Theta}_{1,1}^{*}  - \mathbf{\Theta}_{1,1}^{*} \nonumber\\
&= ( {\mathbf{\Sigma}}_{\mathcal{S}_{1,1}}^{-1} +  \mathbf{M}^*_{1,1} )^{-1} + ( {\mathbf{\Sigma}}_{\mathcal{S}_{1,1}}^{-1} +  \mathbf{M}^*_{1,1} )^{-1} ({\mathbf{\Sigma}}_{\mathcal{S}_{1,1}}^{-1}+\mathbf{M}^*_{1,1}  )\mathbf{\Theta}_{1,1}^{*}  - \mathbf{\Theta}_{1,1}^{*} \label{eq:3prc}\\
&=  ( {\mathbf{\Sigma}}_{\mathcal{S}_{1,1}}^{-1} +  \mathbf{M}^*_{1,1} )^{-1}
\succ \mathbf{0},\nonumber
\end{align}
where (\ref{eq:3pra}) is due to (\ref{eq:en1}) while (\ref{eq:3prb}) and (\ref{eq:3prc}) are due to (\ref{eq:KKT2}).







It can be verified that
\begin{align}
&( \mathbf{\Theta}_{1,1}^{*} + {\mathbf{\Sigma}}_{\mathcal{S}_{1,1}})^{-1} {\mathbf{\Sigma}}_{\mathcal{S}_{1,1}} \nonumber \\
&=( \mathbf{\Theta}_{1,1}^{*} + {\mathbf{\Sigma}}_{\mathcal{S}_{1,1}})^{-1} (\mathbf{\Theta}_{1,1}^{*}+{\mathbf{\Sigma}}_{\mathcal{S}_{1,1}}-\mathbf{\Theta}_{1,1}^{*}) \nonumber \\
&= \mathbf{I}_m - ( \mathbf{\Theta}_{1,1}^{*} + {\mathbf{\Sigma}}_{\mathcal{S}_{1,1}})^{-1} \mathbf{\Theta}_{1,1}^{*}  \nonumber\\
& = \mathbf{I}_m - (( \mathbf{\Theta}_{1,1}^{*} + {\mathbf{\Sigma}}_{\mathcal{S}_{1,1}})^{-1} + \mathbf{M}^*_{1,1}) \mathbf{\Theta}_{1,1}^{*}\label{eq:4dueKKT}  \\
&= \mathbf{I}_m - ( \mathbf{\Theta}_{1,1}^{*} + \tilde{\mathbf{\Sigma}}_{\mathcal{S}_{1,1}})^{-1} \mathbf{\Theta}_{1,1}^{*}  \label{eq:4duedef}\\
&=( \mathbf{\Theta}_{1,1}^{*} + \tilde{\mathbf{\Sigma}}_{\mathcal{S}_{1,1}})^{-1} \tilde{\mathbf{\Sigma}}_{\mathcal{S}_{1,1}},\nonumber
\end{align}
where (\ref{eq:4dueKKT}) is due to \eqref{eq:KKT2}, and (\ref{eq:4duedef}) is due to \eqref{eq:en1}. This proves (\ref{eq:enhance3}).

Note that
\begin{align}
&( \mathbf{\Theta}_{k+1,2i-1}^{*} + {\mathbf{\Sigma}}_{\mathcal{S}_{k+1,2i-1}})^{-1}(\mathbf{\Theta}^*_{k,i}+\mathbf{\Sigma}_{\mathcal{S}_{k+1,2i-1}}) \nonumber \\
&=( \mathbf{\Theta}_{k+1,2i-1}^{*} + {\mathbf{\Sigma}}_{\mathcal{S}_{k+1,2i-1}})^{-1}(\mathbf{\Theta}^*_{k+1,2i-1}+\mathbf{\Sigma}_{\mathcal{S}_{k+1,2i-1}}-(\mathbf{\Theta}^*_{k+1,2i-1}-\mathbf{\Theta}^*_{k,i})) \nonumber \\
&= \mathbf{I}_m - ( \mathbf{\Theta}_{k+1,2i-1}^{*} + {\mathbf{\Sigma}}_{\mathcal{S}_{k+1,2i-1}})^{-1} (\mathbf{\Theta}^*_{k+1,2i-1}-\mathbf{\Theta}^*_{k,i}) \nonumber\\
& = \mathbf{I}_m - (( \mathbf{\Theta}_{k+1,2i-1}^{*} + {\mathbf{\Sigma}}_{\mathcal{S}_{k+1,2i-1}})^{-1} +\mathbf{M}^*_{k+1,2i-1})(\mathbf{\Theta}^*_{k+1,2i-1}-\mathbf{\Theta}^*_{k,i})\label{eq:5dueKKT}  \\
&=\mathbf{I}_m - ( \mathbf{\Theta}_{k+1,2i-1}^{*} + {\tilde{\mathbf{\Sigma}}}_{\mathcal{S}_{k+1,2i-1}})^{-1} (\mathbf{\Theta}^*_{k+1,2i-1}-\mathbf{\Theta}^*_{k,i})  \label{eq:5duedef}\\
&=( \mathbf{\Theta}_{k+1,2i-1}^{*} + {\tilde{\mathbf{\Sigma}}}_{\mathcal{S}_{k+1,2i-1}})^{-1} (\mathbf{\Theta}^*_{k,i}+\tilde{\mathbf{\Sigma}}_{\mathcal{S}_{k+1,2i-1}}),\quad k=1,\cdots,L-2;i=1,\cdots,2^{k-1},\nonumber
\end{align}
where (\ref{eq:5dueKKT}) is due to \eqref{eq:KKT3}, and (\ref{eq:5duedef}) is due to \eqref{eq:en1}. This proves (\ref{eq:enhance4}).  The proof of (\ref{eq:enhance5}) is similar to that of (\ref{eq:enhance4}) and is thus omitted.

Now it remains to prove (\ref{eq:enhance6}) since (\ref{eq:enhance7}) is trivially true. Indeed,
\begin{align}
&( \mathbf{\Sigma}_X + {\mathbf{\Sigma}}_{\mathcal{S}_{L,2i-1}})^{-1}(\mathbf{\Theta}^*_{L-1,i}+\mathbf{\Sigma}_{\mathcal{S}_{L,2i-1}}) \nonumber \\
&=( \mathbf{\Sigma}_X + {\mathbf{\Sigma}}_{\mathcal{S}_{L,2i-1}})^{-1}(\mathbf{\Sigma}_X+\mathbf{\Sigma}_{\mathcal{S}_{L,2i-1}}-(\mathbf{\Sigma}_X-\mathbf{\Theta}^*_{L-1,i})) \nonumber \\
&= \mathbf{I}_m - ( \mathbf{\Sigma}_X + {\mathbf{\Sigma}}_{\mathcal{S}_{L,2i-1}})^{-1} (\mathbf{\Sigma}_X-\mathbf{\Theta}^*_{L-1,i}) \nonumber\\
& = \mathbf{I}_m - (( \mathbf{\Sigma}_X + {\mathbf{\Sigma}}_{\mathcal{S}_{L,2i-1}})^{-1} +\mathbf{M}^*_{L,2i-1})(\mathbf{\Sigma}_X-\mathbf{\Theta}^*_{L-1,i})\label{eq:6dueKKT}  \\
&=\mathbf{I}_m - ( \mathbf{\Sigma}_X + {\tilde{\mathbf{\Sigma}}}_{\mathcal{S}_{L,2i-1}})^{-1} (\mathbf{\Sigma}_X-\mathbf{\Theta}^*_{k,i})  \label{eq:6duedef}\\
&=( \mathbf{\Sigma}_X + {\tilde{\mathbf{\Sigma}}}_{\mathcal{S}_{L,2i-1}})^{-1} (\mathbf{\Theta}^*_{L-1,i}+\tilde{\mathbf{\Sigma}}_{\mathcal{S}_{L,2i-1}}),\quad i=1,\cdots,2^{L-2},\nonumber
\end{align}
where (\ref{eq:6dueKKT}) is due to \eqref{eq:KKT4}, and (\ref{eq:6duedef}) is due to \eqref{eq:en2}.
This completes the proof of Lemma \ref{lemma:enhance}.

\section{Proof of Lemma \ref{lem:sumrate}}\label{app:sumrate}

It can be verified that
\begin{align}
R^*_G(\underline{\mathbf{D}})&=\frac{1}{2} \log\frac{|\mathbf{\Sigma_{X}} + \mathbf{\Sigma}_{\mathcal{S}_{1,1}}|}{|\mathbf{\Sigma}_{\mathcal{S}_{1,1}}|}   + \sum_{k=1}^{L-1} \sum_{i=1}^{2^{k-1}} \frac{1}{2} \log\frac{|  \mathbf{\Theta}^*_{k,i} + \mathbf{\Sigma}_{\mathcal{S}_{k,i}} ||\mathbf{\Sigma}_{X} + \mathbf{\Sigma}_{\mathcal{S}_{k+1,2i-1}}||\mathbf{\Sigma}_{X} + \mathbf{\Sigma}_{\mathcal{S}_{k+1,2i}}|}{|\mathbf{\Sigma}_{X} + \mathbf{\Sigma}_{\mathcal{S}_{k,i}}||  \mathbf{\Theta}^*_{k,i} + \mathbf{\Sigma}_{\mathcal{S}_{k+1,2i-1}} ||  \mathbf{\Theta}^*_{k,i} + \mathbf{\Sigma}_{\mathcal{S}_{k+1,2i}}|}\nonumber\\
&=\frac{1}{2} \log\frac{|  \mathbf{\Theta}^*_{1,1} + \mathbf{\Sigma}_{\mathcal{S}_{1,1}} |}{|\mathbf{\Sigma}_{\mathcal{S}_{1,1}}|}+\sum_{k=2}^{L-1} \sum_{i=1}^{2^{k-1}} \frac{1}{2} \log\frac{|  \mathbf{\Theta}^*_{k,i} + \mathbf{\Sigma}_{\mathcal{S}_{k,i}} |}{|\mathbf{\Sigma}_{X} + \mathbf{\Sigma}_{\mathcal{S}_{k,i}}|}\nonumber\\
&\quad+ \sum_{k=1}^{L-1} \sum_{i=1}^{2^{k-1}} \frac{1}{2} \log\frac{|\mathbf{\Sigma}_{X} + \mathbf{\Sigma}_{\mathcal{S}_{k+1,2i-1}}||\mathbf{\Sigma}_{X} + \mathbf{\Sigma}_{\mathcal{S}_{k+1,2i}}|}{|  \mathbf{\Theta}^*_{k,i} + \mathbf{\Sigma}_{\mathcal{S}_{k+1,2i-1}} ||  \mathbf{\Theta}^*_{k,i} + \mathbf{\Sigma}_{\mathcal{S}_{k+1,2i}}|}\nonumber\nonumber\\
&=\frac{1}{2} \log\frac{|  \mathbf{\Theta}^*_{1,1} + \mathbf{\Sigma}_{\mathcal{S}_{1,1}} |}{|\mathbf{\Sigma}_{\mathcal{S}_{1,1}}|}+\sum_{k=1}^{L-2} \sum_{i=1}^{2^{k-1}} \frac{1}{2} \log\frac{|  \mathbf{\Theta}^*_{k+1,2i-1} + \mathbf{\Sigma}_{\mathcal{S}_{k+1,2i-1}} ||  \mathbf{\Theta}^*_{k+1,2i} + \mathbf{\Sigma}_{\mathcal{S}_{k+1,2i}} |}{|\mathbf{\Sigma}_{X} + \mathbf{\Sigma}_{\mathcal{S}_{k+1,2i-1}}||\mathbf{\Sigma}_{X} + \mathbf{\Sigma}_{\mathcal{S}_{k+1,2i}}|}\nonumber\\
&\quad+ \sum_{k=1}^{L-1} \sum_{i=1}^{2^{k-1}} \frac{1}{2} \log\frac{|\mathbf{\Sigma}_{X} + \mathbf{\Sigma}_{\mathcal{S}_{k+1,2i-1}}||\mathbf{\Sigma}_{X} + \mathbf{\Sigma}_{\mathcal{S}_{k+1,2i}}|}{|  \mathbf{\Theta}^*_{k,i} + \mathbf{\Sigma}_{\mathcal{S}_{k+1,2i-1}} ||  \mathbf{\Theta}^*_{k,i} + \mathbf{\Sigma}_{\mathcal{S}_{k+1,2i}}|}\nonumber\\
&=  \frac{1}{2} \log \frac{|\mathbf{\Theta}^{*}_{1,1} + \mathbf{\Sigma}_{\mathcal{S}_{1,1}}|}{|\mathbf{\Sigma}_{\mathcal{S}_{1,1}}|} + \sum_{k=1}^{L-2} \sum_{i=1}^{2^{k-1}}  \frac{1}{2} \log \frac{| \mathbf{\Theta}^{*}_{k+1,2i-1}+ \mathbf{\Sigma}_{\mathcal{S}_{k+1,2i-1}}|| \mathbf{\Theta}^{*}_{k+1,2i}+ \mathbf{\Sigma}_{\mathcal{S}_{k+1,2i}}|}{| \mathbf{\Theta}^{*}_{k,i}+ \mathbf{\Sigma}_{\mathcal{S}_{k+1,2i-1}}|| \mathbf{\Theta}^{*}_{k,i}+ \mathbf{\Sigma}_{\mathcal{S}_{k+1,2i}}| }    \nonumber \\
&\quad  + \sum_{i=1}^{2^{L-2}} \frac{1}{2} \log \frac{|\mathbf{\Sigma}_{X}+\mathbf{\Sigma}_{\mathcal{S}_{L,2i-1}}||\mathbf{\Sigma}_{X}+\mathbf{\Sigma}_{\mathcal{S}_{L,2i}}|}{|\mathbf{\Theta}^{*}_{L-1,i}+\mathbf{\Sigma}_{\mathcal{S}_{L,2i-1}}||\mathbf{\Theta}^{*}_{L-1,i}+\mathbf{\Sigma}_{\mathcal{S}_{L,2i}}|}\nonumber \\
&=  \frac{1}{2} \log \frac{|\mathbf{\Theta}^{*}_{1,1} + \tilde{\mathbf{\Sigma}}_{\mathcal{S}_{1,1}}|}{|\tilde{\mathbf{\Sigma}}_{\mathcal{S}_{1,1}}|} + \sum_{k=1}^{L-2} \sum_{i=1}^{2^{k-1}}  \frac{1}{2} \log \frac{| \mathbf{\Theta}^{*}_{k+1,2i-1}+ \tilde{\mathbf{\Sigma}}_{\mathcal{S}_{k+1,2i-1}}|| \mathbf{\Theta}^{*}_{k+1,2i}+ \tilde{\mathbf{\Sigma}}_{\mathcal{S}_{k+1,2i}}|}{| \mathbf{\Theta}^{*}_{k,i}+ \tilde{\mathbf{\Sigma}}_{\mathcal{S}_{k+1,2i-1}}|| \mathbf{\Theta}^{*}_{k,i}+ \tilde{\mathbf{\Sigma}}_{\mathcal{S}_{k+1,2i}}| }    \nonumber \\
&\quad  + \sum_{i=1}^{2^{L-2}} \frac{1}{2} \log \frac{|\mathbf{\Sigma}_{X}+\tilde{\mathbf{\Sigma}}_{\mathcal{S}_{L,2i-1}}||\mathbf{\Sigma}_{X}+\tilde{\mathbf{\Sigma}}_{\mathcal{S}_{L,2i}}|}{|\mathbf{\Theta}^{*}_{L-1,i}+\tilde{\mathbf{\Sigma}}_{\mathcal{S}_{L,2i-1}}||\mathbf{\Theta}^{*}_{L-1,i}+\tilde{\mathbf{\Sigma}}_{\mathcal{S}_{L,2i}}|}\label{eq:invokelot}\\
&=  \frac{1}{2} \log \frac{|\mathbf{\Sigma}_X+ \tilde{\mathbf{\Sigma}}_{\mathcal{S}_{1,1}}|}{|\tilde{\mathbf{\Sigma}}_{\mathcal{S}_{1,1}}|}  + \sum_{k=1}^{L-1} \sum_{i=1}^{2^{k-1}}  \frac{1}{2} \log \frac{| \mathbf{\Theta}^{*}_{k,i}+ \tilde{\mathbf{\Sigma}}_{\mathcal{S}_{k,i}}|| \mathbf{\Sigma}_{X}+ \tilde{\mathbf{\Sigma}}_{\mathcal{S}_{k+1,2i-1}}|| \mathbf{\Sigma}_{X}+ \tilde{\mathbf{\Sigma}}_{\mathcal{S}_{k+1,2i}}|}{| \mathbf{\Sigma}_{X}+ \tilde{\mathbf{\Sigma}}_{\mathcal{S}_{k,i}}| | \mathbf{\Theta}^{*}_{k,i}+ \tilde{\mathbf{\Sigma}}_{\mathcal{S}_{k+1,2i-1}}|| \mathbf{\Theta}^{*}_{k,i}+ \tilde{\mathbf{\Sigma}}_{\mathcal{S}_{k+1,2i}}|},\nonumber
\end{align}
where (\ref{eq:invokelot}) is due to (\ref{eq:enhance3})--(\ref{eq:enhance7}).

\section{Proof of Lemma \ref{lemma:GTC}} \label{app:GTC}

It follows by \eqref{eq:enhance8} that
 \begin{align*}
  \tilde{\mathbf{\Sigma}}_{\mathcal{S}_{k+1,2i-1}}-\tilde{\mathbf{\Sigma}}_{\mathcal{S}_{k,i}} ,\tilde{\mathbf{\Sigma}}_{\mathcal{S}_{k+1,2i}}-\tilde{\mathbf{\Sigma}}_{\mathcal{S}_{k,i}}  \succ \mathbf{0}.
  \end{align*}
   Moreover, it is known \cite[Equation (107)]{WV07} that \eqref{eq:KKT_en1} implies
\begin{align}
\mathbf{\Theta}_{k,i}^{*} =& ( \tilde{\mathbf{\Sigma}}_{\mathcal{S}_{k+1,2i-1}}-\tilde{\mathbf{\Sigma}}_{\mathcal{S}_{k,i}}) ^{\frac{1}{2}}(( \tilde{\mathbf{\Sigma}}_{\mathcal{S}_{k+1,2i-1}}-\tilde{\mathbf{\Sigma}}_{\mathcal{S}_{k,i}}) ^{-\frac{1}{2}}(\tilde{\mathbf{\Sigma}}_{\mathcal{S}_{k+1,2i}}-\tilde{\mathbf{\Sigma}}_{\mathcal{S}_{k,i}} ) ( \tilde{\mathbf{\Sigma}}_{\mathcal{S}_{k+1,2i-1}}-\tilde{\mathbf{\Sigma}}_{\mathcal{S}_{k,i}}) ^{-\frac{1}{2}})^{\frac{1}{2}} \nonumber \\
& \quad ( \tilde{\mathbf{\Sigma}}_{\mathcal{S}_{k+1,2i-1}}-\tilde{\mathbf{\Sigma}}_{\mathcal{S}_{k,i}}) ^{\frac{1}{2}} - \tilde{\mathbf{\Sigma}}_{\mathcal{S}_{k,i}} ,\nonumber
\end{align}
which further implies
\begin{align*}
\tilde{\mathbf{\Sigma}}_{\mathcal{S}_{k+1,2i}}-\tilde{\mathbf{\Sigma}}_{\mathcal{S}_{k,i}} = ( \mathbf{\Theta}^*_{k,i}-\tilde{\mathbf{\Sigma}}_{\mathcal{S}_{k,i}}) ( \tilde{\mathbf{\Sigma}}_{\mathcal{S}_{k+1,2i-1}}-\tilde{\mathbf{\Sigma}}_{\mathcal{S}_{k,i}})^{-1}(\mathbf{\Theta}^*_{k,i}-\tilde{\mathbf{\Sigma}}_{\mathcal{S}_{k,i}}).
\end{align*}
Therefore, $\mathbf{\Lambda}_{k,i}$ is indeed a positive semi-definite matrix.

According to \cite[Lemmas 3 and 4]{WV07},
\begin{align}
&\mathbf{\Gamma}_{k,i} \succ \mathbf{0},\nonumber\\
&\tilde{\mathbf{\Sigma}}^{-1}_{\mathcal{S}_{k,i}} =(\mathbf{I}_{m}, \mathbf{I}_{m}) \mathbf{\Gamma}^{-1}_{k,i} (\mathbf{I}_{m}, \mathbf{I}_{m})^{T}.\label{eq:invokewv}
\end{align}
Now set
\begin{align*}
(  \mathbf{H}_{k+1,2i-1},    \mathbf{H}_{k+1,2i} ) = (  \tilde{\mathbf{\Sigma}}_{\mathcal{S}_{k,i}},    \tilde{\mathbf{\Sigma}}_{\mathcal{S}_{k,i}}   ) \mathbf{\Gamma}^{-1}_{k,i}.
\end{align*}
It can be verified that
\begin{align}
\mathbf{H}_{k+1,2i-1}+ \mathbf{H}_{k+1,2i} &= (  \tilde{\mathbf{\Sigma}}_{\mathcal{S}_{k,i}},    \tilde{\mathbf{\Sigma}}_{\mathcal{S}_{k,i}}) \mathbf{\Gamma}^{-1}_{k,i} (\mathbf{I}_{m}, \mathbf{I}_{m})^{T}\nonumber  \\
&= \tilde{\mathbf{\Sigma}}_{\mathcal{S}_{k,i}} ( \mathbf{I}_{m} ,    \mathbf{I}_{m}  )\mathbf{\Gamma}^{-1}_{k,i} (\mathbf{I}_{m}, \mathbf{I}_{m})^{T} \nonumber\\
&=\tilde{\mathbf{\Sigma}}_{\mathcal{S}_{k,i}}\tilde{\mathbf{\Sigma}}^{-1}_{\mathcal{S}_{k,i}}\label{eq:invokeinv}\\
&= \mathbf{I}_{m},\nonumber
\end{align}
where (\ref{eq:invokeinv}) is due to (\ref{eq:invokewv}).
Furthermore,
\begin{align}
&( \mathbf{H}_{k+1,2i-1}, \mathbf{H}_{k+1,2i} ) \mathbf{\Lambda}_{k,i}( \mathbf{H}_{k+1,2i-1}, \mathbf{H}_{k+1,2i})^{T} \nonumber \\
&=( \mathbf{H}_{k+1,2i-1}, \mathbf{H}_{k+1,2i} ) \mathbf{\Gamma}_{k,i}
( \mathbf{H}_{k+1,2i-1}, \mathbf{H}_{k+1,2i})^{T} \nonumber \\
&\quad-( \mathbf{H}_{k+1,2i-1}, \mathbf{H}_{k+1,2i} )\left(
                                                                               \begin{array}{cc}
                                                                                 \tilde{\mathbf{\Sigma}}_{\mathcal{S}_{k,i}} & \tilde{\mathbf{\Sigma}}_{\mathcal{S}_{k,i}} \\
                                                                                 \tilde{\mathbf{\Sigma}}_{\mathcal{S}_{k,i}} & \tilde{\mathbf{\Sigma}}_{\mathcal{S}_{k,i}} \\
                                                                               \end{array}
                                                                             \right)( \mathbf{H}_{k+1,2i-1}, \mathbf{H}_{k+1,2i})^{T}\nonumber\\
&=\tilde{\mathbf{\Sigma}}_{\mathcal{S}_{k,i}} - \tilde{\mathbf{\Sigma}}_{\mathcal{S}_{k,i}} \label{eq:invokeinv2}\\
&= \mathbf{0},\nonumber
\end{align}
where (\ref{eq:invokeinv2}) is due to (\ref{eq:invokewv}).
This completes the proof of Lemma \ref{lemma:GTC}.

\bibliographystyle{IEEEtran}
\bibliography{ref}

\end{document}